\documentclass[aps,prl,twocolumn,superscriptaddress]{revtex4-1}

\usepackage{amsthm}
\usepackage{amsmath,bm}
\usepackage{amssymb}
\usepackage{amsfonts}
\usepackage{graphicx}

\usepackage{txfonts}

\usepackage{float}
\usepackage{quantum} 

\usepackage[colorlinks=true,linkcolor=blue,citecolor=blue,urlcolor=blue]{hyperref}

\newcommand{\eg}{\textit{e.g. }}

\begin{document}

\newcommand{\avg}[1]{\langle#1\rangle}		
\newcommand{\var}{\text{Var}}	
\newcommand{\mse}{\text{MSE}}	
\newcommand{\cov}{\text{Cov}}

\newcommand{\Vs}[1]{\var[#1]}
\newcommand{\MSE}[1]{\mse[#1]}
\newcommand{\cvar}{\var_Q^{B|A}}

\newcommand{\ie}{\textit{i.e.\ }}

\renewcommand{\vec}[1]{\mathbf{#1}}
\newcommand{\bg}[1]{\boldsymbol{#1}}
\newcommand{\inlineheading}[1]{\textit{{#1.---}}}
\newcommand{\mc}[1]{\mathcal{#1}}
\newcommand{\mb}[1]{\mathbf{#1}}
\newcommand{\comment}[1]{{\color{red}[#1]}}
\newcommand{\fisher}{F}
\newcommand{\cfi}{\fisher^{B|A}}
\newcommand{\qfi}{\fisher_Q}
\newcommand{\cqfi}{\fisher_Q^{B|A}}

\newcommand{\assem}{\mc{A}}
\newcommand{\tqfi}{\bar{\mc{F}}}
\newcommand{\tcqfi}{\bar{\mc{F}}^{B|A}}
\newcommand{\steer}{\mc{S}}
\newcommand{\steerAvg}{\steer_\mathrm{avg}}
\newcommand{\steerMax}{\steer_\mathrm{max}}
\newcommand{\pos}[1]{\left[ #1 \right]^+}

\newcommand{\mg}[1]{{\color{magenta}[(MG) #1]}}
\newcommand{\ps}[1]{{\color{red}[PS:  #1]}}
\newcommand{\mf}[1]{{\color{blue}[(MF) #1]}}
\newcommand{\jj}[1]{{\color{cyan}[(JJ) #1]}}
\newcommand{\qh}[1]{{\color{red}[(QH) #1]}}

\title{Detecting Einstein-Podolsky-Rosen steering in non-Gaussian spin states\\from conditional spin-squeezing parameters}

\author{Jiajie Guo}
\affiliation{State Key Laboratory for Mesoscopic Physics, School of Physics, Frontiers Science Center for Nano-optoelectronics, $\&$ Collaborative
Innovation Center of Quantum Matter, Peking University, Beijing 100871, China}

\author{Feng-Xiao Sun}
\affiliation{State Key Laboratory for Mesoscopic Physics, School of Physics, Frontiers Science Center for Nano-optoelectronics, $\&$ Collaborative
Innovation Center of Quantum Matter, Peking University, Beijing 100871, China}

\author{Daoquan Zhu}
\affiliation{State Key Laboratory for Mesoscopic Physics, School of Physics, Frontiers Science Center for Nano-optoelectronics, $\&$ Collaborative
Innovation Center of Quantum Matter, Peking University, Beijing 100871, China}

\author{Manuel Gessner}
\email{manuel.gessner@ens.fr}
\affiliation{Laboratoire Kastler Brossel, ENS-Universit\'{e} PSL, CNRS, Sorbonne Universit\'{e}, Coll\`{e}ge de France, Paris, France}

\author{Qiongyi He}
\email{qiongyihe@pku.edu.cn}
\affiliation{State Key Laboratory for Mesoscopic Physics, School of Physics, Frontiers Science Center for Nano-optoelectronics, $\&$ Collaborative
Innovation Center of Quantum Matter, Peking University, Beijing 100871, China}
\affiliation{Collaborative Innovation Center of Extreme Optics, Shanxi University, Taiyuan, Shanxi 030006, China}
\affiliation{Peking University Yangtze Delta Institute of Optoelectronics, Nantong 226010, Jiangsu, China}

\author{Matteo Fadel}
\email{matteo.fadel@unibas.ch}
\affiliation{Department of Physics, University of Basel, Klingelbergstrasse 82, 4056 Basel, Switzerland}

\date{\today}

\begin{abstract}
We present an experimentally practical method to reveal Einstein-Podolsky-Rosen steering in non-Gaussian spin states by exploiting a connection to quantum metrology. Our criterion is based on the quantum Fisher information, and uses bounds derived from generalized spin-squeezing parameters that involve measurements of higher-order moments. This leads us to introduce the concept of conditional spin-squeezing parameters, which quantify the metrological advantage provided by conditional states, as well as detect the presence of an EPR paradox.
\end{abstract}

\maketitle

\textbf{Introduction.--} Einstein-Podolsky-Rosen (EPR) steering was first termed by Schr\"{o}dinger~\cite{Schrodinger1935} to describe the contradiction to local complementarity in the EPR paradox~\cite{EPR1935}. As an intermediate correlation, EPR steering is stronger than entanglement but not as general as Bell nonlocality~\cite{WisemanPRL2007}. Being easier to generate and detect than nonlocality renders EPR steering a valuable resource for a variety of quantum information tasks~\cite{ReidRMP2009,CavalcantiRPP2017,GuehneRMP2020}, such as quantum teleportation~\cite{QiongyiPRL2015,ChiuNPJQI2016}, one-side device independent QKD~\cite{BranciardPRA2012,GehringNC2015,WalkOpt2016,GallegoPRX2015}, quantum secret sharing~\cite{ArmstrongNP2015,XiangPRA2017,KogiasPRA2017} and assisted quantum metrology~\cite{YadinNC2021}.

\begin{figure}[t]
	\begin{center}
		\includegraphics[width=0.9\columnwidth]{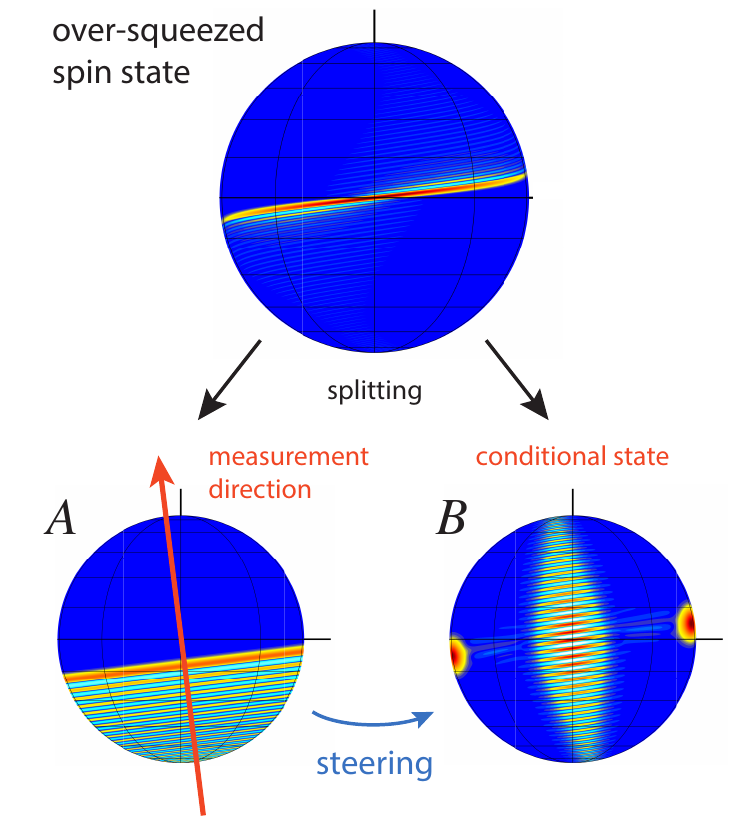}
	\end{center}
	\caption{Illustration of the investigated protocol. A non-Gaussian (over-squeezed) spin state is prepared in an ensemble of particles, that are then distributed to form subsystems $A$ and $B$. Because of quantum correlations, a measurement on $A$ projects $B$ into one of several highly sensitive conditional states. The knowledge of $A$'s measurement setting and result allow $B$ to make the best use of its state by optimising his local measurement. With the criteria we propose, steering between the two subsystems can be concluded.}
	\label{fig:scheme}
\end{figure}

Typically, EPR steering is revealed from the violation of a criterion based on a local uncertainty relation~\cite{ReidPRA1989,ReidRMP2009}. For this reason, such criteria are often expressed in terms of variances of linear operators, and therefore best suited to reveal steering in Gaussian states, where the correlations are fully described by first and second-order moments. Recently, non-Gaussian states were shown to have more competitive advantages in several quantum information protocols~\cite{GuoPRA2019,TakahashiNP2010,LeeNPJQI2019,MariPRL2012}. However, their nontrivial correlations appear in higher-order moments of physical operators, leading to the failure of steering criteria limited to linear observables. To detect non-Gaussian steering, some approaches have taken higher-order moments into account
~\cite{WalbornPRL11,WalbornPRA13}. For example, nonlinear correlations in a three-photon down-conversion process with quadratic steerability index were considered in Ref.~\cite{ShenPRL2015}. A steering criterion derived from Hillery and Zubairy's multimode entanglement criterion~\cite{HilleryPRL2006} has been investigated to detect steering in multipartite scenario~\cite{CavalcantiPRA2011} and further extended to a higher-order version in a two-well BEC ground state~\cite{QiongyiPRA2012}. 
Nevertheless, these methods are specifically tailored to particular states, and a general steering criterion for non-Gaussian states is still highly desirable to further unlock their potential applications.

Nonclassical spin states are many-body quantum states of great interest for fundamental studies as well as for practical applications. For example, squeezed spin states have attracted increasing attention in quantum metrology for precision improvements to overcome the standard quantum limit and are nowadays routinely prepared in a variety of platforms, from solid state systems to atomic ensembles~\cite{PezzeRMP2018}. Recent studies have in particular explored the metrological potential of non-Gaussian spin states, both in theory~\cite{ManuelPRL2019,BaamaraArxiv} and experiment~\cite{StrobelScience2014,BohnetSCIENCE2016,EvrardPRL2019,XuArXiv2021,ColomboArXiv2021}.

Methods derived from quantum metrology~\cite{WinelandPRA1992,SorensenNL2001,SorensenPRL2001,PezzePRL2009,RenPRL2021}, already allow for the efficient detection of multiparticle entanglement without addressing individual spins. In particular, the quantum Fisher information (QFI) constitutes a powerful tool for capturing even strongly non-Gaussian features of quantum states by probing them for their sensitivity under small perturbations~\cite{StrobelScience2014}. 
Very recently, the QFI was also used to formulate a criterion for EPR correlations~\cite{YadinNC2021}, thus providing us with a powerful method for detecting EPR steering in non-Gaussian states.

However, accessing the QFI is often challenging. Determining the QFI of arbitrary mixed states requires full knowledge of the quantum state. On the other hand, efficient approximations based on the full counting statistics demand that a carefully chosen observable is measured with high resolution, which is also difficult in multipartite systems. Spin-squeezing parameters~\cite{WinelandPRA1992} have proven to be efficient alternatives with high practical relevance, especially for Gaussian spin systems; suitable generalizations are also able to capture non-Gaussian features from higher-order moments~\cite{ManuelPRL2019}. But so far they have been limited to the detection of entanglement in a many-spin ensemble with collective measurements.

Here, we introduce the concept of conditional spin-squeezing parameters, and based on that we propose a practical and convenient witness for EPR steering in split nonclassical spin states. For the purpose of detecting non-Gaussian steering, spin observables involving higher-order moments are taken into consideration. An optimization of the measurement within these accessible higher-order observable ensembles leads to conditional nonlinear spin-squeezing parameters, whose potential to detect steering in a wider class of non-Gaussian states is explored. We demonstrate that conditional spin-squeezing parameters approximate the conditional QFI criterion~\cite{YadinNC2021} and as we increase the order of the measured moments this approximation ultimately converges to the QFI criterion. In addition, we also prove that the conditional spin squeezing parameters detect a larger class of steerable correlations than Reid's criterion~\cite{ReidPRA1989,ReidRMP2009}. As a detailed study, we analyze their performance using analytical results for split one-axis-twisted states, where a hierarchy of criteria is clearly shown. Our work provides an experimentally practical tool to witness non-Gaussian steering, which helps to further investigate the quantum information of non-Gaussian spin states and paves a way to exploiting their promising potential.  

\textbf{Phase estimation and the spin-squeezing parameter.--}
In a typical phase estimation protocol, a generator $H$ imprints an unknown parameter $\theta$ on quantum state $\rho$. An observable $M$ is then measured on the probe state $\rho(\theta)$ and an estimator $\theta_{\mathrm{est}}$ for $\theta$ is constructed as a function of the measurement results. This protocol is repeated $m$ times and, for unbiased cases, the variance of the estimator $\var[\theta_\mathrm{est}]$ represents the deviation of the estimate $\theta_\mathrm{est}$ to the parameter $\theta$. A simple estimator known as the method of moments is constructed from the average value of $M$ and yields, in the limit $m\gg 1$ the phase uncertainty $\var[\theta_\mathrm{est}] = \chi^2[\rho,H,M]/m$, where
\begin{equation}
    \chi^2[\rho,H,M] := \dfrac{\var[\rho,M]}{\abs{\langle [H,M] \rangle_{\rho}}^2}
\end{equation}
is the spin-squeezing parameter~\cite{WinelandPRA1992}. 

For an unbiased estimation, a fundamental limit to the sensitivity is given by the Cram$\acute{\text{e}}$r-Rao bound $\var[\theta_\mathrm{est}]\geq(mF[\rho,H,M])^{-1}$, where $F[\rho,H,M]$ is the Fisher information (FI)~\cite{Helstrom1976,Holevo1982}. By optimizing over all observables $M$, the maximum value of the FI defines the quantum Fisher information (QFI), \ie $F_Q[\rho,H]=\max_M F[\rho,H,M]$, which determines the optimal sensitivity potential of the probe state $\rho$~\cite{ParisIJQI2009}. In conclusion, we have $\chi^{-2}[\rho,H,M] \leq F[\rho,H,M] \leq F_Q[\rho,H]$~\cite{BraunsteinPRL1994,PezzePRL2009}. For practical experiments, the achievable sensitivity can be optimized by maximizing $\chi^{-2}[\rho,H,M]$ over a set of measurement operators $M$ that can be realistically implemented~\cite{ManuelPRL2019}. Denoting with $\mathbf{X}$ a basis for such measurements, we can achieve the maximal sensitivity $\max_{M\in\mathrm{span}(\mathbf{X})}\chi^{-2}[\rho,H,M]$.

\textbf{Assisted phase estimation with conditional squeezing parameter.--} In the assisted phase-estimation protocol~\cite{YadinNC2021}, Bob's estimation of $\theta$ is improved by communication from Alice about her measurement setting and result, $Y$ and $b$. This information allows Bob to choose a measurement observable $M\in\mathrm{span}(\mathbf{X})$ that is optimally tailored to the conditional state $\rho^B_{b|Y}$. This way, Bob can achieve on average an estimation sensitivity given by the conditional spin-squeezing parameter~\cite{footnote}
%
\begin{equation}\label{eq:condSSXfixed}
    (\chi^{-2})^{B \vert A}[\assem,H,\mathbf{X},Y]  := \sum_b p(b|Y) \max_{M\in\mathrm{span}(\mathbf{X})}\chi^{-2}[\rho^{B}_{b|Y},H,M] \;.
\end{equation}
Here, we introduced the definition of assemblages $\assem(b,Y) = p(b|Y) \rho_{b|Y}^B$, which are determined by the local probability distribution $p(b|Y)$ for results $b$ conditioned on Alice's measurement observable $Y$ and Bob's conditional state $\rho_{b|Y}^B$. 
Note that the ultimate limit in phase estimation for a specific measurement $Y$ for Alice is expressed by the conditional Fisher information~\cite{YadinNC2021}
\begin{equation}\label{eq:conditionalQFIX}
    \cfi[\assem,H,Y] := \sum_b p(b|Y) \qfi[\rho^B_{b|Y},H] \;,
\end{equation}
In fact, we can obtain a chain of inequalities (see Supplementary Information Sec.~IA)
\begin{equation}\label{eq:lwrbndmethods}
\cfi[\assem,H,Y] \geq  \, (\chi^{-2})^{B \vert A}[\assem,H,\mathbf{X},Y] \geq  \frac{ \abs{\expect{[H,M]}_{\rho^{B}}}^2 }{\var^{B|A}[\assem,M,Y]},
\end{equation}
which results in a hierarchy of EPR steering criteria that will be discussed below.

\textbf{Connection to EPR steering.--} Besides the estimation of the phase $\theta$, we could be interested in estimating its generator $H$. As $\theta$ and $H$ are conjugate variables, the uncertainty principle prevents their simultaneous knowledge with arbitrary precision~\cite{BraunsteinAnnPhys1996}. By making use of EPR steering from Alice to Bob in the assisted metrology protocol, an inference of these properties can be realized below the local uncertainty limit, which is given by $\Vs{\theta_{\text{est}}}\Vs{H_{\text{est}}} \geq (4m)^{-1}$~\cite{YadinNC2021}. Here, $\var[\theta_{\text{est}}]$ and $\var[H_{\text{est}}]$ are inference variances. Based on the additional information of Alice's measurement setting $X$ and result $a$, Bob uses the estimator $h_{\text{est}}(a)$ to predict the result $h$ of his local measurement $H$ with the inference variance $\var[H_{\text{est}}]:=\sum_{a,h} p(a,h|X,H)(h_{\text{est}}(a)-h)^2$. A lower bound for $\Vs{H_\mathrm{est}}$ is given by the conditional variance~\cite{ReidRMP2009}
\begin{equation}\label{eq:cvar}
    \var^{B|A}[\assem,H,X] := \sum_a p(a|X) \var[\rho^B_{a|X},H] \;.
\end{equation}
Analogously, we can obtain $\var[\theta_{\text{est}}]$ with a different choice for Alice's measurement settings (denoted by $Y$) and Bob's measurement $M$. The uncertainty bound is a witness of EPR steering, whose violation implies that Alice's measurements can steer Bob's states to overcome Bob's local phase-generator uncertainty relation.

Consider now a local hidden state (LHS) model~\cite{WisemanPRL2007}, described by a classical random variable $\lambda$ with probability distribution $p(\lambda)$. The assemblage in this case can be written as $\assem(a,X)=\sum_{a,\lambda} p(a|X,\lambda) p(\lambda) \sigma_{\lambda}^B$. In Ref.~\cite{YadinNC2021}, a steering criterion based on the QFI is proposed. For any LHS model, it holds
\begin{equation}\label{eq:main}
    \Delta_1 :=  \cfi[\assem,H,Y]-4\var^{B|A}[\assem,H,X]   \leq 0 \;,
\end{equation}
independently of the choices of Alice's measurement. 

As the conditional squeezing parameter provides a lower bound to the conditional FI~(\ref{eq:lwrbndmethods}), this allows us to formulate the following steering criterion: For any assemblage $\assem$ that admits a LHS model, we have
\begin{equation}\label{eq:condSSLHScondition}
    \Delta_2 := (\chi^{-2})^{B \vert A}[\assem,H,\mathbf{X},Y]-4 \var^{B|A}[\assem,H,X]   \leq 0 \;.
\end{equation}
This criterion is one of the main results of this work. The violation of (\ref{eq:condSSLHScondition}) reveals useful EPR steering in the assisted metrological protocol.

\textbf{Reduction to Reid's criterion.--}
From the previous chain of inequalities~(\ref{eq:lwrbndmethods}), it is immediate to recover Reid's criterion~\cite{ReidPRA1989,ReidRMP2009} (here expressed in a linearized form)
\begin{equation}\label{eq:reidopt_XY}
    \Delta_3 := \dfrac{|\langle [H,M]\rangle_{\rho^B}|^2}{\var^{B|A}[\assem,M,Y]}-4 \var^{B|A}[\assem,H,X] \leq 0 \;.
\end{equation}
As defined in Eq.~(\ref{eq:cvar}), the conditional variance represents the average of individual variances for Bob's conditional states, determined as the minimized inference variance~\cite{ReidRMP2009}. For linear observables, Reid's criterion is very powerful for Gaussian states; in a continuous variable setting, it can be shown to be necessary and sufficient for steering detection by Gaussian measurements~\cite{WisemanPRL2007}, while it may fail to detect steering in non-Gaussian cases. Because of the right-hand side in Eq.~(\ref{eq:lwrbndmethods}), we obtain $\Delta_2 \geq \Delta_3$, that means the criterion $\Delta_2$ we proposed contains all LHS models based on Reid's uncertainty relations. A crucial advantage of $\Delta_2$ over Reid's criterion $\Delta_3$ is the possibility to adapt the measurement observable $M$ to each conditional state $\rho_{b|Y}$ individually [see the maximization in Eq.~(\ref{eq:condSSXfixed})], while in~(\ref{eq:reidopt_XY}) only a single $M$ is used for the entire assemblage. As we will see below, this leads in particular to an increased potential to reveal non-Gaussian EPR steering in a wider class of states, especially when the set $\mathbf{X}$ contains higher-order moments of the collective spins.

\textbf{Reduction to linear-estimate Reid's criterion.--}
If Bob's estimator $h_\mathrm{est}(a)$ depends linearly on Alice's measurement result $a$ and takes the form $h_\mathrm{est}(a)=ga+d$, optimal estimates are obtained by minimizing the inference variance $\Vs{H_\mathrm{est}}$. Based on that, a well-known linear-estimate Reid's criterion commonly used in experiments is
\begin{align}\label{Reid}
\Delta_4 := \dfrac{|\langle [H,M]\rangle_{\rho^B}|^2}{\Vs{Y+g' M}}-4\Vs{X-g H}  \leq 0 \;.
\end{align}
Note that in a Gaussian system, where quantum correlations are well characterized with first and second-order moments, the best estimator equals the optimized linear estimator~\cite{ReidRMP2009}. In this case, we have $\Vs{X-g H}=\var^{B|A}[\assem,H,X]$, which leads to $\Delta_3=\Delta_4$ for Gaussian states.

We define the maximum value of the left-hand side of the above criteria as $\delta_i = \max_{H\in\mathrm{span}(\mathbf{H}),M\in\mathrm{span}(\mathbf{X})} \Delta_i$, respectively (see Supplementary Information Sec.~IB,IIB,IIIC,IIID). As a result, a hierarchy of criteria reads
\begin{align}\label{eq:hierarchy1}
\delta_1 \geq \delta_2 \geq \delta_3 \geq \delta_4 \;.
\end{align}
In the following, we will compare these criteria for a scenario of experimental relevance.

\begin{figure*}[t]
	\begin{center}
		\includegraphics[width=\textwidth]{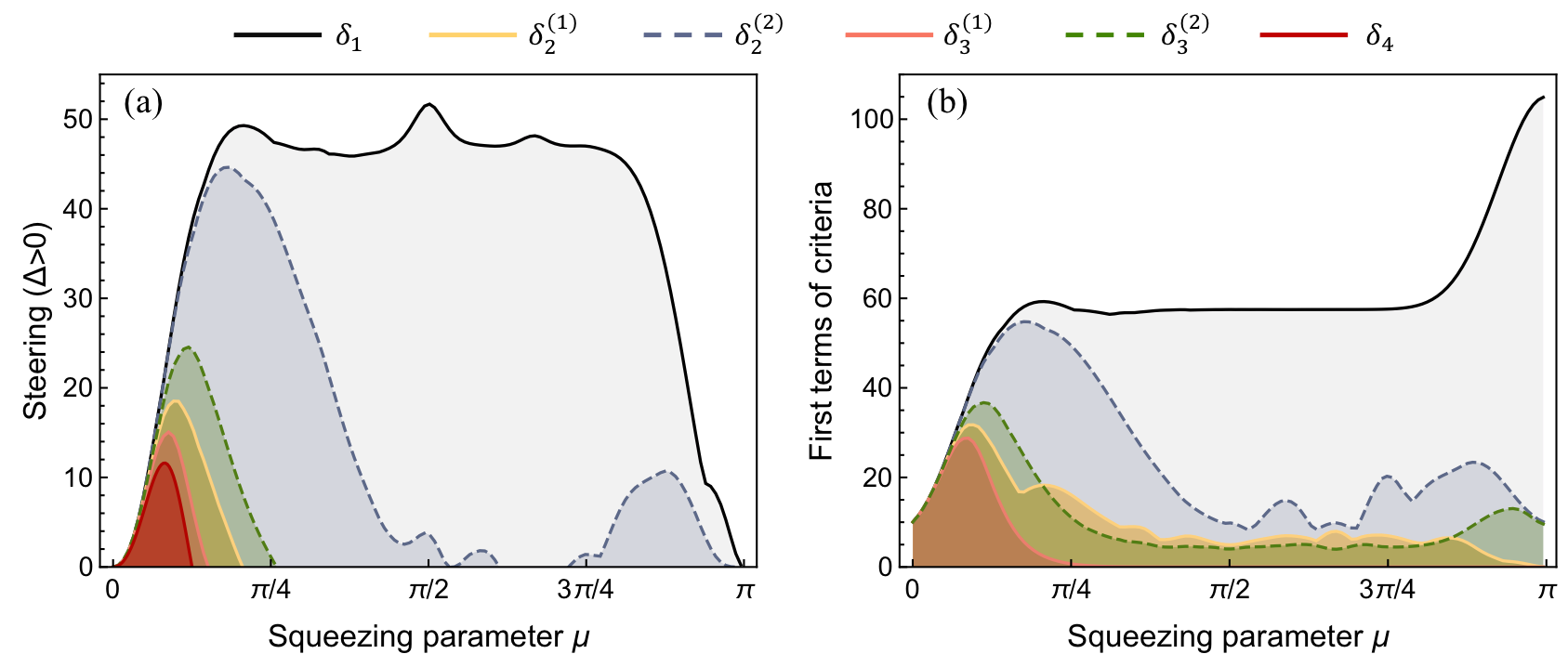}
	\end{center}
	\caption{Steering detection for spilt spin states $\ket{\Phi(\mu)}$ with total atomic number N=20. \textbf{(a)} A hierarchy of criteria with optimized measurement operators, where the dashed lines represent the optimized second-order criteria $\delta_2^{(2)}$ and $\delta_3^{(2)}$ involving nonlinear spin operators ensemble $\textbf{S}^{(2)}$. \textbf{(b)} Comparison among first terms of optimized criteria $\delta_{1,2,3}$, which is also the chain of inequalities in~(\ref{eq:lwrbndmethods}).}
	\label{fig:mainresults}
\end{figure*}

\textbf{Split spin-squeezed state.--} Squeezed states play a key role in measurement sensitivity enhancement, overcoming the standard quantum limit in quantum metrology~\cite{CavesPRD1981,WinelandPRA1992,PezzeRMP2018,TothJPA2014,MaPR2011}. Experimentally, these are routinely adopted in optical and atom interferometers, \eg Ramsey spectroscopy~\cite{PezzeRMP2018}, atom clocks~\cite{LudlowRMP2015} and gravitational-wave detection~\cite{GodaNP2008}. Many experiments have demonstrated the preparation of spin squeezing in atomic systems~\cite{RiedelNL2010,BohnetSCIENCE2016,LerouxPRL2010,MonzPRL2011,ChalopinNC2018}. Here, we focus on spin squeezed states prepared via one-axis twisting (OAT) dynamics~\cite{KitagawaPRA1993}, one of the paramount approaches to generate squeezing via atomic collisions.

Initially, the atomic ensemble is in a coherent spin state, which consists of $N$ spins polarized along the $x$ direction. Then, the time evoution of the OAT Hamiltonian $H=\hbar\chi S_z^2$ can be parametrized by $\mu=2\chi t$, and the state reads
\begin{equation}
\ket{\psi(\mu)} = \frac{1}{\sqrt{2^N}} \sum_{k=0}^N \sqrt{{N}\choose{k}} \; e^{- i \frac{\mu}{2} (N/2 - k)^2} \ket{k}\;.
\end{equation} 
Here, $k$ labels the basis of Dicke states $\ket{k}$. For short interaction times, OAT results in nearly-Gaussian spin squeezed states, however, as time increases the state becomes over-squeezed and significantly non-Gaussian~\cite{PezzePRL2009}.

In order to use such states for assisted phase-estimation protocols, we spatially separate each of the two spin up/down modes into two parts, $A$ and $B$, which can be described by a beam splitter transformation, so that a four-mode split spin-squeezed state is finally obtained~\cite{YumangNJP2019}. This state has also been prepared experimentally~\cite{FadelScience2018}.  In the bipartite Dicke basis, it can be written as
\begin{align}
\ket{\Phi(\mu)} = \frac 1 {2^{N}} \sum_{N_A=0}^{N}&\sum_{k_A=0}^{N_A}\sum_{k_B=0}^{N-N_A}  \sqrt{\binom{N}{N_A} \binom{N_A}{k_A} \binom{N-N_A}{k_B}} \notag\\ &\quad\times e^{- i \frac{\mu}{2} (N/2 - k_A - k_B)^2} \ket{k_A}_{N_A} \ket{k_B}_{N-N_A} \;.
\end{align}
where $N_\alpha$ is the number of particles in $\alpha\in\{A,B\}$, $\ket{k_\alpha}_{N_\alpha}$ represents the Dicke states with $k_\alpha$ spins down and $N_\alpha-k_\alpha$ spins up along the $z$ direction.

The properties of split spin states can be characterized by local collective spin observables, that is $\mathbf{S}^{\alpha} = \sum_{i\in\alpha} \boldsymbol{\sigma}^{(i)}/2$, where $\boldsymbol{\sigma}^{(i)}$ is the vector of Pauli matrices acting on particle $i$.

\textbf{Measurement optimisation.--} The sharpest formulations of the above criteria are obtained by optimizing the measurement observables $X,Y$ for Alice, and $H,M$ for Bob, respectively. In linear cases, $\textbf{S}^{(1)}=(S_x,S_y,S_z)$ is used to describe all the local collective spin operators for Alice and Bob. The linear measurements that appear in the first-order criterion $\Delta_2^{(1)}$ have been extensively used in experiments~\cite{PezzeRMP2018}. We consider the squeezing direction of the split spin squeezed states to define the $\textbf{z}$ axis and the anti-squeezing direction $\textbf{y}$. Therefore, Alice's measurement settings $X$ and $Y$ can be also restricted to the $yz$ plane. To the optimize measurement directions of Alice's and Bob's observables, we construct the moment matrix~\cite{ManuelPRL2019} and covariance matrix for Bob's conditional states (see Supplementary Information Sec.~IIB for details).

For the metrological characterization of non-Gaussian spin states, higher-order moments of physical observables are of great importance~\cite{ManuelPRL2019}. Note that
in~\cite{ShenPRL2015}, the linear-estimate Reid's criterion, i.e., $\Delta_4$, with non-linear operators is introduced. Here, we also extend our conditional spin-squeezing parameter based criterion $\Delta_2$ and general Reid's criterion $\Delta_3$ to nonlinear version, where they can be further optimized by taking into account a set $\mathbf{X}$ of higher-order measurements for Bob (i.e. $M$). When $M$ is a product of up to $n$ linear spin observables, optimized nonlinear $\delta^{(n)}_{2,3}$ are still upper bounded by the Fisher criterion $\delta_1$, but this bound becomes increasingly tight as $n$ grows larger. To be concrete, let us start with the second-order criteria, $\mathbf{X}=\textbf{S}^{(2)}$, where an ensemble of linear and symmetric quadratic spin operators  $\textbf{S}^{(2)}=(S_x,S_y,S_z,S_x^2,S_y^2,S_z^2,\frac{1}{2}\{S_x ,S_y\}, \frac{1}{2}\{S_x ,S_z\},\frac{1}{2}\{S_y ,S_z\})$ is introduced. We refer to Ref.~\cite{XuArXiv2021} for an experimental measurement of such observables. Moreover, it was shown in Ref.~\cite{BaamaraArxiv} how such observables may become accessible by a second OAT evolution before the measurement of a linear spin observable. For our second-order criteria $\Delta_2^{(2)}$ and $\Delta_3^{(2)}$, Alice's measurements $X,Y$ and Bob's generator $H$ for the phase imprinting evolution are still linear, but the measurement operator for Bob takes into account second-order operators $M=\textbf{m} \cdot \textbf{S}^{(2)}$, with $\textbf{m} \in \mathbb{R}^{9}$. Analogously, we can also obtain higher-order criteria, leading to another chain of inequalities 
\begin{align}\label{eq:hierarchy2}
    \delta_i^{(1)} \leq \delta_i^{(2)}\leq \delta_i^{(3)}\leq \dots \leq \delta_1\;,
\end{align}
for both criteria $i=2,3$.

As illustrated in Fig.~\ref{fig:mainresults}(a), for $N=20$ split spin states, we obtain analytically optimized criteria $\delta_{i}$ evolved with OAT squeezing parameter $\mu$, where both the hierarchy relations~(\ref{eq:hierarchy1}) and~(\ref{eq:hierarchy2}) are shown clearly. At small squeezing levels $\mu$, the evolution generates near-Gaussian split spin-squeezed states and all criteria detect steering and tend to converge. However, if only linear operators are considered, Reid's criteria $\delta_4$ and $\delta_3^{(1)}$ decay soon in the non-Gaussian area for longer evolution times, while the conditional spin-squeezing parameter criterion $\delta_2^{(1)}$ reveals steering in a wider range of states. Furthermore, when it comes to nonlinear version, the second-order $\delta_2^{(2)}$ shows great advantages. The Fisher criterion $\delta_1$ bounds all other criteria from above during the entire dynamics, but the conditional spin parameter criterion $\delta_2$ is more practical experimentally and it will tend towards the Fisher criterion as higher-order measurement operators are involved. All of these criteria compare the phase estimation sensitivity [first term in Eqs.~(\ref{eq:main}), (\ref{eq:condSSLHScondition}), (\ref{eq:reidopt_XY}) and~(\ref{Reid})] to the estimation variance for the generator (second term). While the second term hardly varies between the criteria, the hierarchy can be traced back to the chain of inequalities~(\ref{eq:lwrbndmethods}), which is reflected in Fig.~\ref{fig:mainresults}(b).

\textbf{Conclusions.--} We have proposed a non-Gaussian steering criterion based on conditional spin squeezing parameters. By introducing nonlinear operators and optimizing measurement within accessible higher-order observables, the criterion shows an improved ability to reveal EPR steering with a larger range of non-Gaussian states. This approach is related by a hierarchy to other criteria such as Reid's criteria and metrological complementarity based on the QFI. The key advantage of metrology-based steering criteria is the ability to adjust the measurement observable to each conditional state individually. Our steering criterion is experimentally feasible and constitutes a general method to reveal non-Gaussian steering in a bipartite scenario. This work provides a powerful approach to further investigate nonlinear EPR correlations in non-Gaussian states and takes a step further to unlock the promising applications for non-Gaussian systems.

\textbf{Acknowledgments.--}
\begin{acknowledgments}
	This work is supported by the National Natural Science Foundation of China (Grants No. 11975026, No. 61675007, and No. 12004011), Beijing Natural Science Foundation (Grant No. Z190005), the Key R$\&$D Program of Guangdong Province (Grant No. 2018B030329001), and the LabEx ENS-ICFP: ANR-10-LABX-0010 / ANR-10-IDEX-0001-02 PSL*.
\end{acknowledgments}



\clearpage
\newpage

\begin{widetext}

\section{Supplementary information}

\section{I.\quad  Fisher criterion}\label{si:section1}

\subsection{A.\quad  Definition: Fisher criterion}

As proposed in~\cite{YadinNC2021}, a formulation of an EPR criterion can be derived from the \textit{quantum conditional variance}  
\begin{equation}\label{eq:optcondvar}
	\cvar[\assem,H] := \min_X \, \sum_a p(a|X) \var[\rho^B_{a|X},H] \;.
\end{equation}
and the \textit{quantum conditional Fisher information}
\begin{align} \label{eq:assisted_qfi}
	\cqfi[\assem,H] 	& := \max_Y \, \sum_b p(b|Y) \qfi[\rho^B_{b|Y},H] \;.
\end{align}
For any assemblage $\assem$ that admits a LHS model, the following bound holds:
\begin{align}\label{eq:fishercriterion}
	\cqfi[\assem,H] \leq 4 \cvar[\assem,H] \;.
\end{align}

Also here, note that the quantum conditional variance (\ref{eq:optcondvar}) and the quantum conditional Fisher information (\ref{eq:assisted_qfi}) are defined as an optimisation over Alice's measurement. For simplicity, however, we always compute a conditional Fisher Information by choosing a specific measurement $Y$ for Alice
\begin{equation}\label{eq:conditionalQFIX}
    \cfi[\assem,H,Y] := \sum_b p(b|Y) \qfi[\rho^B_{b|Y},H] \;,
\end{equation}
and also identify the conditional variance as calculated based on a specific measurement $X$ for Alice with the symbol
\begin{equation}\label{eq:conditionalvar}
    \var^{B|A}[\assem,H,X] := \sum_a p(a|X) \var[\rho^B_{a|X},H] \;,
\end{equation}
In summary, we then have $\cqfi[\assem,H] \geq  \cfi[\assem,H,Y]$ and $\cvar[\assem,H] \leq \var^{B|A}[\assem,H,X]$. So after choosing specific measurement setting for Alice, the Fisher criterion~(\ref{eq:fishercriterion}) can be written as:
\begin{equation}
   \Delta_1:= \cfi[\assem,H,Y] - 4\var^{B|A}[\assem,H,X] \leq 0 \;,
\end{equation} 
its violation indicates EPR steering from Alice to Bob.

\subsection{B.\quad Calculation: Fisher criterion}
Note that in our case, the split spin state is a pure global state and therefore the conditional states $\rho^B_{a|X}$ of Bob will be pure, too, i.e., $\rho^B_{a|X} = \ket{\phi^B_{a|X}}\bra{\phi^B_{a|X}}$. Using $F_Q[|\psi\rangle,H]=4\var[|\psi\rangle,H]$~\cite{BraunsteinPRL1994} (for simplicity we write $F_Q[|\psi\rangle,H]$ instead of $F_Q[|\psi\rangle\langle \psi|,H]$ for pure states and similarly for the variance), we obtain from~(\ref{eq:conditionalQFIX}) that
\begin{equation}
    \cfi[\assem,H,Y] = 4 \sum_b p(b|Y) \var[\ket{\phi^B_{a|X}},H] = 4 \var^{B|A}[\assem,H,Y] \;.
\end{equation}
Decomposing the Hamiltonian $H$ in terms of a vector of observables $\mathbf{H}$ as $H=\mathbf{n}^T\mathbf{H}$ with real coefficient vector $\mathbf{n}$, the conditional variance can be written as
\begin{align}\label{eq:condvariance}
\var^{B|A}[\assem,H,Y]=\mathbf{n}^T\boldsymbol{\Gamma}^{B|A}[\assem,\mathbf{H},Y]\mathbf{n} \;,
\end{align}
where we defined the conditional covariance matrices by introducing
\begin{align}
    \boldsymbol{\Gamma}[\rho,\mathbf{H}]_{ij} &= \frac{1}{2}\langle H_i H_j+H_j H_i\rangle_{\rho}-\langle H_i\rangle_{\rho}\langle H_j\rangle_{\rho} \;,\label{eq:covariancematrix}\\
    \boldsymbol{\Gamma}^{B|A}[\assem,\mathbf{H},Y]&=\sum_b p(b|Y)\boldsymbol{\Gamma}[\rho^B_{b|Y},\mathbf{H}]\;.
\end{align}
And analogously, we also have $\var^{B|A}[\assem,H,X]=\mathbf{n}^T\boldsymbol{\Gamma}^{B|A}[\assem,\mathbf{H},X]\mathbf{n}$, which allows us to optimize the witness over the Hamiltonian $H$ analytically:
\begin{align}
    \max_{H\in\mathrm{span}(\mathbf{H})}\Delta_1&=4\max_{\mathbf{n}}\left(\var^{B|A}[\assem,H,Y]-\var^{B|A}[\assem,H,X]\right)\notag\\
    &=\lambda_{\max} (\boldsymbol{\Gamma}^{B|A}[\assem,\mathbf{H},Y]-\boldsymbol{\Gamma}^{B|A}[\assem,\mathbf{H},X]).
\end{align}

Since the squeezing direction lies in the yz-plane, we also restrict Alice's measurement direction to this plane, and represent it by the angle $\phi$ on the yz-plane: $\cos(\phi)\textbf{y}+\sin(\phi)\textbf{z}$. We label the measurement direction $\phi_X$ for $X$ , $\phi_Y$ for $Y$. After optimizing also over Alice's measurements, we thus obtain the optimized witness
\begin{align}\label{eq:delta1}
\delta_1:=4\max_{\phi_X,\phi_Y} \, \lambda_{\max} (\boldsymbol{\Gamma}^{B|A}[\assem,\mathbf{H},Y]-\boldsymbol{\Gamma}^{B|A}[\assem,\mathbf{H},X]) \;.
\end{align}
For the comparison of the hierarchy~(\ref{eq:lwrbndmethods}), we limit the optimization to the first term, i.e., $4\max_{\phi_Y}\lambda_{\max} (\boldsymbol{\Gamma}^{B|A}[\assem,\mathbf{H},Y])$.

\subsection{C.\quad A chain of inequalities}

It holds for arbitrary measurements $M$ that~\cite{PezzePRL2009}
\begin{align}
\qfi[\rho,H] \geq \frac{\abs{\expect{[H,M]}_{\rho}}^2}{\var[\rho,M]} \;.
\end{align}
In analogy to results in~\cite{YadinNC2021}, based on the Cauchy-Schwarz inequality and the definitions of the conditional Fisher information and the conditional variance, we obtain the chain of inequalities
\begin{align}\label{eq:chaininequalities}
\cfi[\assem,H,Y]
& \geq  \, \sum_b p(b|Y) \frac{\abs{\expect{[H,M]}_{\rho^{B}_{b|Y}}}^2}{\var[\rho^{B}_{b|Y},M]} \nonumber \\
& \geq  \, \frac{\abs{ \sum_b p(b|Y) \expect{[H,M]}_{\rho^{B}_{b|Y}} }^2}{ \sum_b p(b|Y) \var[\rho^{B}_{b|Y},M]} \nonumber \\
& = \frac{ \abs{\expect{[H,M]}_{\rho^{B}}}^2 }{\sum_b p(b|Y) \var[\rho^{B}_{b|Y},M]} \nonumber \\
& = \frac{ \abs{\expect{[H,M]}_{\rho^{B}}}^2 }{\var^{B|A}[\assem,M,Y]}.
\end{align}
We will use this result to lower bound the Fisher criterion when using specific measurements for Alice. Note that this relation (\ref{eq:chaininequalities}) leads to the hierarchy of criteria.


\section{II.\quad Conditional spin-squeezing parameter based criterion}\label{si:section2}

\subsection{A.\quad Definition: Conditional spin-squeezing parameter based criterion}

Rather than using the QFI, which corresponds to the maximal sensitivity achievable with an optimal estimation strategy and observable, we can quantify Bob's sensitivity assuming the estimation of the phase parameter with the method of moments based on the data obtained from a specific collective spin observable. In this case, we arrive at the quantum conditional spin-squeezing parameter (where we consider the same measurement $M$ for different results of Alice, even if in full generality one can have $M_{a|X}$)
\begin{equation}\label{eq:squeezingparameter}
    (\chi^{-2}_Q)^{B \vert A}[\assem,H,\mathbf{X}]  := \max_Y \, \sum_b p(b|Y) \max_{M\in\mathrm{span(\mathbf{X})}} \, \chi^{-2}[\rho^{B}_{b|Y},H,M] \;,
\end{equation}
where the (generalized) spin-squeezing parameter for rotations generated by $H$, and measurements of $M$, is defined as
\begin{equation}
    \chi^{2}[\rho,H,M] := \dfrac{\var[\rho,M]}{\abs{\langle [H,M] \rangle_{\rho}}^2} \;.
\end{equation}
The bound $\qfi[\rho,H] \geq \chi^{-2}[\rho,H,M]$ for all $M$ implies that 
\begin{align}
    \qfi^{B \vert A}[\assem, H] \geq (\chi^{-2}_Q)^{B \vert A}[\assem,H,\mathbf{X}] \;.
\end{align}
This can be used to introduce a steering criterion weaker than the Fisher criterion but in general stronger than the Reid's criterion, namely
\begin{equation}
    (\chi^{-2}_Q)^{B \vert A}[\assem,H,\mathbf{X}] \leq 4 \var^{B|A}_Q[\assem,H] \;.
\end{equation}

For a specific choice of the measurement $Y$, we define conditional spin-squeezing parameters
\begin{equation}\label{eq:condSSXfixed_SUPP}
    (\chi^{-2})^{B \vert A}[\assem,H,\mathbf{X},Y]  := \sum_b p(b|Y) \max_{M\in\mathrm{span}(\mathbf{X})} \, \chi^{-2}[\rho^{B}_{b|Y},H,M] \;,
\end{equation}
such that, according to the first line in Eq.~\eqref{eq:chaininequalities} we have
\begin{align}
    \qfi^{B \vert A}[\assem, H,Y] \geq (\chi^{-2})^{B \vert A}[\assem,H,\mathbf{X},Y] \;.
\end{align}

This can be used to introduce the conditional spin-squeezing parameter based steering criterion 
\begin{equation}\label{eq:Delta2}
    \Delta_2:=(\chi^{-2})^{B \vert A}[\assem,H,\mathbf{X},Y] - 4 \var^{B|A}[\assem,H,X] \leq 0 \;.
\end{equation}
Furthermore, the criterion $\Delta_2$ can be extended to a higher-order version $\Delta_2^{(n)}$, if nonlinear measurement settings $\textbf{X}^{(n)}$ are taken into consideration
\begin{equation}\label{eq:highorderDelta2}
    \Delta_2^{(n)}:=(\chi^{-2})^{B \vert A}[\assem,H,\mathbf{X}^{(n)},Y] - 4 \var^{B|A}[\assem,H,X] \leq 0 \;.
\end{equation}
With higher-order measurement settings, $\Delta_2^{(n)}$ will get closer to the Fisher criterion $\Delta_1$, and will be eventually equal to $\Delta_1$ when $\mathbf{X}$ becomes a complete basis of operators~\cite{ManuelPRL2019}.

\subsection{B.\quad Calculation: Conditional spin-squeezing parameter based criterion}

Expanding $H$ and $M$ in terms of vectors of observables as $H=\mathbf{n}^T\mathbf{H}$ and $M=\mathbf{m}^T\mathbf{X}$, the first term of (\ref{eq:Delta2}) can be written as
\begin{align}
    \chi^{-2}[\rho^{B}_{b|Y},H,M]
    =\frac{\abs{\langle [H,M] \rangle_{\rho^B_{b|Y}}}^2}{\var[\rho^B_{b|Y},M]} =\frac{|\mathbf{n}^T\mathbf{C}[\rho^B_{b|Y},\mathbf{H},\mathbf{X}]\mathbf{m}|^2}{\mathbf{m}^T\boldsymbol{\Gamma}[\rho^B_{b|Y},\mathbf{X}]\mathbf{m}} \;,
\end{align}
where $\textbf{C}[\rho,\mathbf{H},\mathbf{X}]$ is the commutator matrix with elements $(\mathbf{C}[\rho,\mathbf{H},\mathbf{X}])_{ij}=-i\langle [X_j,H_i]\rangle_{\rho}$, and $\boldsymbol{\Gamma}[\rho,\mathbf{X}]$ is the covariance matrix similar to (\ref{eq:covariancematrix}) with elements $(\boldsymbol{\Gamma}[\rho,\mathbf{X}])_{ij}=\mathrm{Cov}(X_i,X_j)_{\rho}$. Let us first discuss the optimization over $M$ in Eq.~(\ref{eq:condSSXfixed_SUPP}). As introduced in~\cite{ManuelPRL2019}, we have
\begin{align}\label{eq:maxcondchi}
    \max_{M\in\mathrm{span}(\mathbf{X})} \, \chi^{-2}[\rho^{B}_{b|Y},H,M]
    =\max_{M\in\mathrm{span}(\mathbf{X})} \, \dfrac{\abs{\langle [H,M] \rangle_{\rho^B_{b|Y}}}^2}{\var[\rho^B_{b|Y},M]}
    =\textbf{n}^{T}\textbf{M}[\rho^{B}_{b|Y},\textbf{H},\mathbf{X}]\textbf{n} \;,
\end{align}
Here, the moment matrix $\textbf{M}[\rho,\mathbf{H},\mathbf{X}]$ for an arbitrary state $\rho$ is defined as
\begin{align}\label{eq:momentstandard}
   \mathbf{M}[\rho,\mathbf{H},\mathbf{X}]:=\textbf{C}^T[\rho,\mathbf{H},\mathbf{X}] \boldsymbol{\Gamma}^{-1}[\rho,\mathbf{X}] \mathbf{C}[\rho,\mathbf{H},\mathbf{X}] \;,
\end{align}

For the conditional spin squeezing parameter, we obtain
\begin{align}
    (\chi^{-2})^{B \vert A}[\assem,H,\mathbf{X},Y]
    &=\sum_b p(b|Y) \left(\textbf{n}^T \textbf{M}[\rho^{B}_{b|Y},\mathbf{H},\mathbf{X}] \textbf{n}\right) \notag \\
    &=\textbf{n}^T \textbf{M}^{B|A}[\assem,\mathbf{H},\mathbf{X},Y]\textbf{n} \;,
\end{align}
where we introduced the conditional moment matrix
\begin{align}\label{eq:momentconditional}
    \textbf{M}^{B|A}[\assem,\mathbf{H},\mathbf{X},Y]=\sum_b p(b|Y)\textbf{M}[\rho^{B}_{b|Y},\mathbf{H},\mathbf{X}] \;.
\end{align} 
Next, we choose an optimal Hamiltonian $H$, i.e., the vector $\mathbf{n}$ to maximize the steering witness. Together with (\ref{eq:condvariance}), we obtain
\begin{align}
     \max_{H\in\mathrm{span}(\mathbf{H}),M\in\mathrm{span}(\mathbf{X})} \, \left( (\chi^{-2})^{B \vert A}[\assem,H,\mathbf{X},Y] - 4 \var^{B|A}[\assem,H,X] \right)
     &=\max_{\mathbf{n}} \left( \textbf{n}^T \textbf{M}^{B|A}[\assem,\mathbf{H},\mathbf{X},Y]\textbf{n}-4 \mathbf{n}^T\boldsymbol{\Gamma}^{B|A}[\assem,\mathbf{H},X]\mathbf{n} \right)\notag\\
     &=\lambda_{\max}(\textbf{M}^{B|A}[\assem,\mathbf{H},\mathbf{X},Y]-4 \boldsymbol{\Gamma}^{B|A}[\assem,\mathbf{H},X]) \;.
\end{align}
At last we choose optimal measurements for Alice. In analogy to our previous derivation we obtain
\begin{align}\label{eq:delta2}
    \delta_2:=\max_{\phi_X,\phi_Y} \, \lambda_{\max}(\textbf{M}^{B|A}[\assem,\mathbf{H},\mathbf{X},Y]-4\boldsymbol{\Gamma}^{B|A}[\assem,\mathbf{H},X]) \;.
\end{align}
For Alice, the optimal measurement direction $\phi_X,\phi_Y$ can be found by searching in the yz-plane ($\phi_{X,Y}\in[0,\pi]$). And for Bob, the optimal phase generator's direction $\mathbf{n}_\mathrm{opt}$ is the eigenvector corresponding to the maximum eigenvalue in (\ref{eq:delta2}). Finally, the optimal measurement direction $\mathbf{m}_\mathrm{opt}$ for the other measurement operator $M$ is $\mathbf{m}_\mathrm{opt}=\alpha \boldsymbol{\Gamma}^{-1}[\rho^{B}_{b|Y},\mathbf{X}] \mathbf{C}[\rho^{B}_{b|Y},\mathbf{H},\mathbf{X}]\mathbf{n}_\mathrm{opt}$, with a real normalization constant $\alpha$~\cite{ManuelPRL2019}. 

Note that here we optimize $M$ for each conditional state $\rho^B_{b|Y}$ individually, i.e., Bob may choose to measure a different observable, depending on the information that is provided by Alice. Below when we introduce Reid's criterion, we will provide a different optimization, where only one optimal $M$ may be chosen for one Alice's measurement $Y$. Besides the hierarchy~(\ref{eq:chaininequalities}), this is the main advantage of $\Delta_2$ compared to Reid's criterion $\Delta_3$.

We obtain a higher-order versions~(\ref{eq:highorderDelta2}) by replacing Bob's linear measurement setting $\mathbf{X}$ with nonlinear one $\mathbf{X}^{(n)}$
\begin{align}\label{eq:highorderdelta2}
    \delta_2^{(n)}:=\max_{\phi_X,\phi_Y} \, \lambda_{\max}(\textbf{M}^{B|A}[\assem,\mathbf{H},\mathbf{X}^{(n)},Y]-4\boldsymbol{\Gamma}^{B|A}[\assem,\mathbf{H},X]) \;.
\end{align}
When only the linear collective spin operators are involved $\textbf{X}^{(1)}= \textbf{S}^{(1)}$, the first term of $\delta_2$ is $\max_{\phi_Y}\lambda_{\max}(\textbf{M}^{B|A}[\assem,\mathbf{H},\mathbf{X}^{(1)},Y]$. And for n-order collective spin operators $\textbf{X}^{(n)}= \textbf{S}^{(n)} (n=2,3,4,...)$, it becomes $\max_{\phi_Y}\lambda_{\max}(\textbf{M}^{B|A}[\assem,\mathbf{H},\mathbf{X}^{(n)},Y]$.
%

\section{III.\quad Reid's criterion}\label{si:section3}

\subsection{A.\quad Definition: Reid's criterion}

In Reid's criterion~\cite{ReidRMP2009}, based on the information of Alice's measurement $X$ and result $a$, Bob chooses an estimator $h_\mathrm{est}(a)$ to predict the value and measures $H$ to yield the result $h$. The average deviation is called inference variance 
\begin{align}\label{eq:inferencevar}
\var[H_\mathrm{est}]:=\sum_{a,h}p(a,h|X,H)(h_\mathrm{est}(a)-h)^2 \;,
\end{align}
where $p(a,h|X,H)$ is the joint probability. The other inference variance $\var[M_\mathrm{est}]$ can be obtained analogously by measurement settings $Y$ and $M$.

The sharpest formulation of Eq.~(\ref{eq:inferencevar}) is thus obtained by minimizing the estimation error. The optimal estimator $h_{\mathrm{est}}(a)=\mathrm{Tr}\{\rho^B_{a|X}H\}$ attains the lower bound   $\Vs{H_\mathrm{est}} \geq  \sum_{a} p(a|X) \var[\rho^B_{a|X},H]$, where $\var[\rho,H]=\langle H^2\rangle_{\rho}-\langle H\rangle_{\rho}^2$ is the variance with $\langle O\rangle_{\rho}=\mathrm{Tr}\{\rho O\}$.
Optimising over Alice's measurement setting $X$ leads to the quantum conditional variance $\cvar[\assem,H]$ (\ref{eq:optcondvar}), and any fixed choice of $X$ yields its upper bound $\var^{B|A}[\assem,H,X]$~(\ref{eq:conditionalvar}).
In summary, we then have (for any $H_\mathrm{est}$)
\begin{align}\label{eq:Varhierarchy}
\cvar[\assem,H] \leq \var^{B|A}[\assem,H,X] \leq \Vs{H_\mathrm{est}} \;.
\end{align}
Based on the quantum conditional variance, Reid's criterion for an EPR paradox consists of a violation of the local uncertainty limit
\begin{equation}\label{Reidopt}
    \cvar[\assem,H]\cvar[\assem,M]\geq \dfrac{ |\langle [H,M]\rangle_{\rho^B}|^2}{4} \;.
\end{equation}
If we chose specific measurement settings $X$, $Y$ for Alice, we have
\begin{equation}\label{eq:reidopt_XY_SUPP}
    \var^{B|A}[\assem,H,X] \var^{B|A}[\assem,M,Y] \geq \dfrac{ |\langle [H,M]\rangle_{\rho^B}|^2}{4} \;,
\end{equation}
this can be written in the linearised form as
\begin{equation}\label{eq:Delta3}
    \Delta_3 := \dfrac{ |\langle [H,M]\rangle_{\rho^B}|^2}{\var^{B|A}[\assem,M,Y]}-4\var^{B|A}[\assem,H,X]  \leq 0 \;.
\end{equation}
The inequality $\Delta_3\leq 0$ holds for all non-steerable assemblages, i.e., those that admit an LHS model, thus steering is detected by a violation of this inequality.

In the simplest situations, in particular for Gaussian states, it is optimal to consider a linear estimator, i.e., $h_\mathrm{est}(a)=ga+d$, This yields an extensively used linear-estimate Reid's criterion~\cite{ReidPRA1989}
\begin{equation}\label{eq:Delta4}
    \Delta_4 := \dfrac{ |\langle [H,M]\rangle_{\rho^B}|^2}{\var[Y+g'M]}-4\var[X-gH]  \leq 0 \;.
\end{equation}

\subsection{B.\quad Discussion on estimators and variances in Reid's criterion.} 
In this section, we will distinguish between MSE (mean squared error) and the variance. We consider the remote estimation of measurement results $h$ performed on Bob's system from the results $a$ that are acquired from local measurements on Alice's system. For each estimator $h_{\mathrm{est}}(a)$ we define the mean squared error
\begin{align}
    \MSE{H_\mathrm{est}}  := \sum_{a,h} p(a,h|X,H) \left( h_{\mathrm{est}}(a) - h \right)^2 \;,
\end{align}
and the variance
\begin{align}
    \Vs{H_\mathrm{est}}  := \sum_{a,h} p(a,h|X,H) \left( h_{\mathrm{est}}(a) - \langle h_{\mathrm{est}}\rangle \right)^2 \;,
\end{align}
where $\langle h_{\mathrm{est}}\rangle=\sum_{a}p(a|X)h_{\mathrm{est}}(a)$ is the average of the estimator. For unbiased estimators, we have $\langle h_{\mathrm{est}}\rangle=h$ and $\MSE{H_\mathrm{est}}=\Vs{H_\mathrm{est}}$. In general we have the relation
\begin{align}
    \MSE{H_\mathrm{est}} &= \Vs{H_\mathrm{est}} + \langle H_{\mathrm{est}}\rangle^2-2\langle H_{\mathrm{est}}H\rangle+\langle H^2\rangle \;.
\end{align}

Let us now focus on a linear estimator constructed from the eigenvalues $a$ of Alice's measurement observable $X$ for the estimation of the result $h$ of Bob's measurement of $H$. We thus take $h_{\mathrm{est}}(a)=ga$, where $g\in\mathbb{R}$ is a constant. We obtain 
\begin{align}
    \MSE{H_\mathrm{est}}  = \langle ( gX - H )^2\rangle_{\rho^{AB}} \;,
\end{align}
and
\begin{align}
    \Vs{H_\mathrm{est}} &= \sum_{a,h} p(a,h|X,H) \left( ga - g\langle X\rangle \right)^2=\Vs{\rho^A,gX} \;,
\end{align}
where $\langle h_{\mathrm{est}}\rangle = g\langle X\rangle$.

It is clear that
\begin{align}
    \Vs{\rho^{AB},gX-H}\leq \MSE{H_\mathrm{est}} \;,
\end{align}
with equality when $g\langle X\rangle = \langle H\rangle$.

Specifically, we focus on the observables $H=S^B_y$ and $M=S^B_z$ and we denote by $y^{A/B}$ and $z^{A/B}$ the measurement results, i.e., the eigenvalues of $S^{A/B}_y$ and $S^{A/B}_z$. We estimate Bob's result $z^B$ for $S^B_z$ from Alice's result $z^A$ for $S^A_z$ using the linear estimator $z^B_{\mathrm{est}}(z^A)=-g_z z^A$. This will lead to
\begin{align}
    \MSE{H_{\mathrm{est}}}=\langle (g_zS_z^A+S_z^B)^2\rangle_{\rho^{AB}} \;,
\end{align}
and if $\langle g_zS_z^A+S_z^B\rangle_{\rho^{AB}}=0$, i.e., if the estimator is unbiased, this coincides with $\Vs{\rho^{AB},g_zS_z^A+S_z^B}$. 

In our cases, since the squeezing direction is on the yz-plane, the spin expectation for reduced states is
\begin{align}
    \langle S_y \rangle_{\rho^{A(B)}}=\langle S_z \rangle_{\rho^{A(B)}}=0 \;.
\end{align}
so we have
\begin{align}
    \langle g_z S_z^A+S_z^B\rangle_{\rho^{AB}}=g_z \langle S_z^A \rangle_{\rho^A}+\langle S_z^B \rangle_{\rho^B}=0 \;.
\end{align}
Similarly, we also have $\langle g_y S_y^A+S_y^B\rangle_{\rho^{AB}}=0$. So it is proved that the estimator is unbiased in our case.

\subsection{C.\quad Calculation: general Reid's criteiron.}

The general Reid's criterion $\Delta_3$ is defined in Eq.~(\ref{eq:Delta3}), and its first term can be written as
\begin{align}
    \frac{\abs{\expect{[H,M]}_{\rho^B}}^2 }{\var^{B|A}[\assem,M,Y]}&=\frac{|\mathbf{n}^T\mathbf{C}[\rho^B,\mathbf{H},\mathbf{X}]\mathbf{m}|^2}{\sum_bp(b|Y)\mathbf{m}^T\boldsymbol{\Gamma}[\rho^B_{b|Y},\mathbf{X}]\mathbf{m}}\notag\\
    &=\frac{|\mathbf{n}^T\mathbf{C}[\rho^B,\mathbf{H},\mathbf{X}]\mathbf{m}|^2}{\mathbf{m}^T\boldsymbol{\Gamma}^{B|A}[\assem,\mathbf{X},Y]\mathbf{m}} \;,
\end{align}
where  $\mathbf{C}[\rho^B,\mathbf{H},\mathbf{X}]$ is the commutator matrix for Bob's reduced states and $\boldsymbol{\Gamma}^{B|A}[\assem,\mathbf{X},Y]$ is the conditional covariance variance.
In analogy to Eq.~(\ref{eq:maxcondchi}), the maximization over $M$ is given by
\begin{align}
\max_{M\in\mathrm{span}(\mathbf{X})} \, \frac{\abs{\expect{[H,M]}_{\rho^B}}^2 }{\var^{B|A}[\assem,M,Y]}=\textbf{n}^{T} \textbf{M}[\assem,\mathbf{H},\mathbf{X},Y] \textbf{n} \;,
\end{align} 
where
\begin{equation}\label{eq:momentOReid}
    \textbf{M}[\assem,\mathbf{H},\mathbf{X},Y]=\textbf{C}[\rho^B,\mathbf{H},\mathbf{X}]^{T} {\boldsymbol{\Gamma}^{B|A}[\assem,\mathbf{X},Y]}^{-1} \textbf{C}[\rho^B,\mathbf{H},\mathbf{X}] \;.
\end{equation}
Note that this optimization yields a single optimal measurement observable $M$, which is different from Eq.~(\ref{eq:delta2}). 

With Eq.~(\ref{eq:condvariance}), we then optimize the generator $H$
\begin{align}
 \max_{H\in\mathrm{span}(\mathbf{H}),M\in\mathrm{span}(\mathbf{X})}\left(\frac{\abs{\expect{[H,M]}_{\rho^B}}^2 }{\var^{B|A}[\assem,M,Y]}-4\var^{B|A}[\assem,H,X]\right)
 &=\max_{\mathbf{n}}\left(\mathbf{n}^T\textbf{M}[\assem,\mathbf{H},\mathbf{X},Y]\mathbf{n}-4\mathbf{n}^T\boldsymbol{\Gamma}^{B|A}[\assem,\mathbf{H},X]\mathbf{n}\right)\notag\\
 &=\lambda_{\max}(\textbf{M}[\assem,\mathbf{H},\mathbf{X},Y]-4\boldsymbol{\Gamma}^{B|A}[\assem,\mathbf{H},X]) \;.
\end{align}
After also optimizing over Alice's measurement, we finally have
\begin{align}\label{eq:delta3}
\delta_3&:=\max_{\phi_X,\phi_Y} \, \lambda_{\max}(\textbf{M}[\assem,\mathbf{H},\mathbf{X},Y]-4\boldsymbol{\Gamma}^{B|A}[\assem,\mathbf{H},X]) \;.
\end{align}
Here, Bob's optimal direction $\mathbf{n}_\mathrm{opt}$ is still the eigenvector corresponding to the maximum eigenvalue in Eq.~(\ref{eq:delta3}), and the  optimal direction for Bob's measurement $M$ is $\mathbf{m}_\mathrm{opt}=\beta {\boldsymbol{\Gamma}^{B|A}[\assem,\mathbf{X},Y]}^{-1} \textbf{C}[\rho^B,\mathbf{H},\mathbf{X}] \mathbf{n}_\mathrm{opt}$, with a real normalization constant $\beta$.

By introducing higher-order measurement settings $\mathbf{X}^{(n)}$ for Bob, we can obtain a nonlinear version of Reid's criterion 
\begin{align}\label{eq:highorderdelta3}
\delta_3^{(n)} := \max_{\phi_X,\phi_Y} \, \lambda_{\max}(\textbf{M}[\assem,\mathbf{H},\mathbf{X}^{(n)},Y]-4\boldsymbol{\Gamma}^{B|A}[\assem,\mathbf{H},X]) \;.
\end{align}
The first term is $\max_{\phi_Y}\lambda_{\max}(\textbf{M}[\assem,\mathbf{H},\mathbf{X},Y])$ for linear version $\delta_3$, and is $\max_{\phi_Y}\lambda_{\max}(\textbf{M}[\assem,\mathbf{H},\mathbf{X}^{(n)},Y])$ for higher-order $\delta_3^{(n)}$.

\subsection{D.\quad Calculation: linear-estimate Reid's criterion}

Linear-estimate Reid's criterion is in Eq.~(\ref{eq:Delta4}), where the real constants $g$ and $g'$ are chosen to minimize the inferred variances, respectively. In our split spin state cases, we consider $X$ and $H$ as local collective spin operators in the anti-squeezing direction $S_y^{A,B}$, and $Y$ and $M$ are local collective spin operators in squeezing direction $S_z^{A,B}$, and the constants $g, g'$ are replaced with $g_y, g_z$, thus we obtain
\begin{align}
    &\frac{\partial \var[g_y S_{y}^A-S_{y}^B]_{\rho^{AB}}}{\partial g_y}=0 \qquad\rightarrow\qquad g_{y(\mathrm{opt})} = \dfrac{{\langle S_y^A S_y^B\rangle}_{\rho^{AB}}-{\langle S_y^A\rangle}_{\rho^{A}} {\langle S_y^B\rangle}_{\rho^{B}}}{\langle {S_y^A}^2\rangle_{\rho^{A}}-{\langle S_y^A\rangle}^2_{\rho^{A}}} \;, \\
    &\frac{\partial \var[g_z S_{z}^A+S_{z}^B]_{\rho^{AB}}}{\partial g_z}=0 \qquad\rightarrow\qquad g_{z(\mathrm{opt})} = \dfrac{{\langle S_z^A\rangle}_{\rho^{A}} {\langle S_z^B\rangle}_{\rho^{B}}-{\langle S_z^A S_z^B\rangle}_{\rho^{AB}}}{\langle {S_z^A}^2\rangle_{\rho^{A}}-{\langle S_z^A\rangle}^2_{\rho^{A}}} \;.
\end{align}
where $\rho^{AB}$ is the split spin state, $\rho^{A(B)}$ is the reduced state for Alice(Bob).

Thus, the optimized linear-estimate Reid's criterion for the split spin states is
\begin{equation}\label{eq:delta4}
    \delta_4 := \dfrac{ |\langle S_x^B\rangle_{\rho^B}|^2}{\var[g_{z(\mathrm{opt})} S_{z}^A+S_{z}^B]_{\rho^{AB}}}-4\var[g_{y(\mathrm{opt})} S_{y}^A-S_{y}^B]_{\rho^{AB}}  \;.
\end{equation}
%

\section{IV.\quad Split Spin States } \label{si:section4}

Spin squeezed states (SSS) of an ensemble of $N$ particles can be prepared from a coherent spin state polarized along the $x$ direction, through the action of the one-axis twisting (OAT) Hamiltonian $H=\hbar\chi S_z^2$. The resulting state can be parametrized by $\mu=2\chi t$, where $t$ is the interaction time, and it takes the form
\begin{equation}
\ket{\psi(\mu)} = \dfrac{1}{\sqrt{2^N}} \sum_{k=0}^N \sqrt{{N}\choose{k}} e^{- i \frac{\mu}{2} (N/2 - k)^2} \ket{k} \;.
\end{equation} 
As before, the states $\ket{k}$ represent the Dicke states with $k$ spins down and $N-k$ spins up along $z$. In other words, $k$ represent the number of excitations in two bosonic modes.

Consider now the effect of applying a beam splitter transformation to the two modes $a$ and $b$, which results in a split spin states~\cite{YumangNJP2019}, which has been created experimentally~\cite{FadelScience2018}. In the bipartite Dicke basis, this state can be written as
\begin{align}
\ket{\Phi(\mu)} = \frac 1 {2^{N}} \sum_{N_A=0}^{N}\sum_{k_A=0}^{N_A}\sum_{k_B=0}^{N-N_A}  \sqrt{\binom{N}{N_A} \binom{N_A}{k_A} \binom{N-N_A}{k_B}} e^{- i \frac{\mu}{2} (N/2 - k_A - k_B)^2} \ket{k_A}_{N_A} \ket{k_B}_{N-N_A} \;.
\end{align}

\subsection{A.\quad EPR criterion}
Let us consider a measurement of $S_{\vec{n}}^A$ by Alice, which yields the results $N_A$ (the number of particles in Alice side) and $l_A$ (the result of the spin measurement along the chosen direction). Let $\ket{l_A}_{\vec{n},N_A}$ denote the corresponding eigenstates of $S_{\vec{n}}^A$ in the spin-$N_A/2$ subspace, such that we can write
\begin{align}
S_{\vec{n}}^A=\sum_{N_A=0}^N\sum_{l_A=0}^{N_A}\left(l_A-\frac{N_A}{2}\right)\ket{l_A}_{\mathbf{n},N_A}\bra{l_A}_{\mathbf{n},N_A}.
\end{align}
The event of obtaining measurement results $(l_A,N_A)$ for Alice occurs with probability 
\begin{align}
p(l_A,N_A|S_{\vec{n}}^A)=\frac 1 {2^{2N}} \sum_{k_A=0}^{N_A}\sum_{k'_A=0}^{N_A}\sum_{k_B=0}^{N-N_A}\binom{N}{N_A}\binom{N-N_A}{k_B}\sqrt{\binom{N_A}{k_A}\binom{N_A}{k'_A}} e^{- i \frac{\mu}{2} (N/2 - k_A - k_B)^2}e^{i \frac{\mu}{2} (N/2 - k'_A - k_B)^2} \bra{l_A}_{\vec{n},N_A}\ket{k_A}_{N_A}\bra{k'_A}_{N_A}\ket{l_A}_{\vec{n},N_A}
\end{align}
when $0\leq l_A\leq N_A$ and zero otherwise, 
and produces the conditional states  
\begin{align}
\ket{\Phi(\mu)^B}_{l_A,N_A|S_{\vec{n}}^A}=\frac {p(l_A,N_A|S_{\vec{n}}^A)^{-1/2}} {2^{N}} \sum_{k_A=0}^{N_A}\sum_{k_B=0}^{N-N_A}\sqrt{\binom{N}{N_A}\binom{N_A}{k_A}\binom{N-N_A}{k_B}} e^{- i \frac{\mu}{2} (N/2 - k_A - k_B)^2} \bra{l_A}_{\vec{n},N_A}\ket{k_A}_{N_A} \ket{k_B}_{N-N_A}.
\end{align}
for Bob's system.

Let us consider a phase shift generated on Bob's subsystem by the observable $S_{\vec{m}}^B$. Writing
\begin{align}
S_{\vec{m}}^B=\sum_{N_A=0}^N\sum_{l_B=0}^{N-N_A}\left(l_B-\frac{N-N_A}{2}\right)\ket{l_B}_{\mathbf{m},N-N_A}\bra{l_B}_{\mathbf{m},N-N_A},
\end{align}
we obtain the following first and second moments for the conditional states 
\begin{align}
\langle S_{\vec{m}}^B\rangle_{l_A,N_A}&=\sum_{l_B=0}^{N-N_A}\left(l_B-\frac{N-N_A}{2}\right)|\bra{l_B}_{\vec{m},N-N_A}\ket{\Phi(\mu)^B}_{l_A,N_A|S_{\vec{n}}^A}|^2,\\
\langle (S_{\vec{m}}^B)^2\rangle_{l_A,N_A}&=\sum_{l_B=0}^{N-N_A}\left(l_B-\frac{N-N_A}{2}\right)^2|\bra{l_B}_{\vec{m},N-N_A}\ket{\Phi(\mu)^B}_{l_A,N_A|S_{\vec{n}}^A}|^2,
\end{align}
with
\begin{align}
\bra{l_B}_{\vec{m},N-N_A}\ket{\Phi(\mu)^B}_{l_A,N_A|S_{\vec{n}}^A}=\frac {p(l_A,N_A|S_{\vec{n}}^A)^{-1/2}} {2^{N}} \sum_{k_A=0}^{N_A}\sum_{k_B=0}^{N-N_A}\sqrt{\binom{N}{N_A}\binom{N_A}{k_A}\binom{N-N_A}{k_B}} e^{- i \frac{\mu}{2} (N/2 - k_A - k_B)^2} \bra{l_A}_{\vec{n},N_A}\ket{k_A}_{N_A} \bra{l_B}_{\vec{m},N-N_A}\ket{k_B}_{N-N_A}.
\end{align}
This allows us to obtain the variance of the conditional states
\begin{align}
\var[\ket{\Phi(\mu)^B}_{l_A,N_A|S_{\vec{n}}^A},S_{\vec{m}}^B]=\langle (S_{\vec{m}}^B)^2\rangle_{l_A,N_A}-\langle S_{\vec{m}}^B\rangle_{l_A,N_A}^2,
\end{align}
and an upper bound for the quantum conditional variance
\begin{align}
\cvar[\ket{\Phi(\mu)},S_{\vec{m}}^B]\leq \sum_{N_A=0}^N\sum_{l_A=0}^{N_A}p(l_A,N_A|S_{\vec{n}}^A)\var[\ket{\Phi(\mu)^B}_{l_A,N_A|S_{\vec{n}}^A},S_{\vec{m}}^B].
\end{align}
Since the conditional states are pure, we have
\begin{align}
\qfi[\ket{\Phi(\mu)^B}_{l_A,N_A|S_{\vec{n}}^A},S_{\vec{m}}^B]=4\var[\ket{\Phi(\mu)^B}_{l_A,N_A|S_{\vec{n}}^A},S_{\vec{m}}^B]
\end{align}
and we immediately obtain also a lower bound for the conditional quantum Fisher information as
\begin{align}
\cqfi[\ket{\Phi(\mu)},S_{\vec{m}}^B]\geq 4\sum_{N_A=0}^N\sum_{l_A=0}^{N_A}p(l_A,N_A|S_{\vec{n}}^A)\var[\ket{\Phi(\mu)^B}_{l_A,N_A|S_{\vec{n}}^A},S_{\vec{m}}^B].
\end{align}

\subsection{B.\quad Reduced quantum Fisher information and variance}
The reduced state for Bob's system reads 
\begin{align}
\rho^B= \frac 1 {2^{2N}} \sum_{N_A=0}^{N}\sum_{k_A=0}^{N_A}\sum_{k_B=0}^{N-N_A}\sum_{k'_B=0}^{N-N_A}\binom{N}{N_A} \binom{N_A}{k_A} \sqrt{\binom{N-N_A}{k_B}\binom{N-N_A}{k'_B}} e^{- i \frac{\mu}{2} (N/2 - k_A - k_B)^2}  e^{i \frac{\mu}{2} (N/2 - k_A - k'_B)^2} \ket{k_B}_{N-N_A}  \bra{k'_B}_{N-N_A}.
\end{align}
This yields first and second moments as 
\begin{align}
\langle S_{\vec{m}}^B\rangle_{\rho^B}=\frac 1 {2^{2N}} \sum_{N_A=0}^{N}\sum_{l_B=0}^{N-N_A}\sum_{k_A=0}^{N_A}\sum_{k_B=0}^{N-N_A}\sum_{k'_B=0}^{N-N_A}&\binom{N}{N_A} \binom{N_A}{k_A} \sqrt{\binom{N-N_A}{k_B}\binom{N-N_A}{k'_B}} e^{- i \frac{\mu}{2} (N/2 - k_A - k_B)^2}  e^{i \frac{\mu}{2} (N/2 - k_A - k'_B)^2}\notag\\&\quad\times\left(l_B-\frac{N-N_A}{2}\right)\bra{l_B}_{\mathbf{m},N-N_A}
 \ket{k_B}_{N-N_A}  \bra{k'_B}_{N-N_A}\ket{l_B}_{\mathbf{m},N-N_A}
 \end{align}
and 
\begin{align}
\langle (S_{\vec{m}}^B)^2\rangle_{\rho^B}=\frac 1 {2^{2N}} \sum_{N_A=0}^{N}\sum_{l_B=0}^{N-N_A}\sum_{k_A=0}^{N_A}\sum_{k_B=0}^{N-N_A}\sum_{k'_B=0}^{N-N_A}&\binom{N}{N_A} \binom{N_A}{k_A} \sqrt{\binom{N-N_A}{k_B}\binom{N-N_A}{k'_B}} e^{- i \frac{\mu}{2} (N/2 - k_A - k_B)^2}  e^{i \frac{\mu}{2} (N/2 - k_A - k'_B)^2}\notag\\&\quad\times\left(l_B-\frac{N-N_A}{2}\right)^2\bra{l_B}_{\mathbf{m},N-N_A}
 \ket{k_B}_{N-N_A}  \bra{k'_B}_{N-N_A}\ket{l_B}_{\mathbf{m},N-N_A},
 \end{align}
leading to the variance
\begin{align}
\var[\rho^B,S_{\vec{m}}^B]=\langle (S_{\vec{m}}^B)^2\rangle_{\rho^B}-\langle S_{\vec{m}}^B\rangle_{\rho^B}^2.
\end{align}

To determine the quantum Fisher information, we must diagonalize the state $\rho^B$:
\begin{align}
\rho^B&=\frac 1 {2^{2N}} \sum_{N_A=0}^{N}\sum_{k_A=0}^{N_A}\sum_{k_B=0}^{N-N_A}\sum_{k'_B=0}^{N-N_A}\binom{N}{N_A} \binom{N_A}{k_A} \sqrt{\binom{N-N_A}{k_B}\binom{N-N_A}{k'_B}} e^{- i \frac{\mu}{2} (N/2 - k_A - k_B)^2}  e^{i \frac{\mu}{2} (N/2 - k_A - k'_B)^2} \ket{k_B}_{N-N_A}  \bra{k'_B}_{N-N_A}\notag\\
&=\frac{1}{2^{N}}\sum_{N_A=0}^{N}\sum_{k_A=0}^{N_A}\frac{1}{2^{N_A}}\binom{N}{N_A} \binom{N_A}{k_A}\ket{\Psi(N_A,k_A)}\bra{\Psi(N_A,k_A)}\notag\\
&=\sum_{N_A=0}^{N}\sum_{k_A=0}^{N_A}p(N_A,k_A)\ket{\Psi(N_A,k_A)}\bra{\Psi(N_A,k_A)},
\end{align}
where the eigenstates read
\begin{align}
\ket{\Psi(N_A,k_A)}=\frac{1}{\sqrt{2^{N-N_A}}}\sum_{k_B=0}^{N-N_A}\sqrt{\binom{N-N_A}{k_B}}e^{- i \frac{\mu}{2} (N/2 - k_A - k_B)^2}\ket{k_B}_{N-N_A},
\end{align}
and the corresponding eigenvalues are
\begin{align}
p(N_A,k_A)=\frac{1}{2^{N+N_A}}\binom{N}{N_A} \binom{N_A}{k_A}.
\end{align}
We can now use the expression $\qfi[\rho^B,S_z^B]=2\sum_{i,j}\frac{(p_i-p_{j})^2}{p_i+p_{j}}|\langle\psi_i|S_z^B|\psi_j\rangle|^2$ to write
\begin{align}
\qfi[\rho^B,S_{\vec{m}}^B]&=2\sum_{N_A=0}^{N}\sum_{k_A,k_A'=0}^{N_A}\frac{(p(N_A,k_A)-p(N_A,k'_A))^2}{p(N_A,k_A)+p(N_A,k'_A)}\left|\bra{\Psi(N_A,k_A)}S_{\vec{m}}^B\ket{\Psi(N_A,k'_A)}\right|^2\notag\\
&=2\sum_{N_A=0}^{N}\sum_{l_B=0}^{N-N_A}\sum_{k_A,k_A'=0}^{N_A}\frac{(p(N_A,k_A)-p(N_A,k'_A))^2}{p(N_A,k_A)+p(N_A,k'_A)}\left(l_B-\frac{N-N_A}{2}\right)^2|\braket{\Psi(N_A,k_A)}{l_B}_{\mathbf{m},N-N_A}|^2|\braket{\Psi(N_A,k'_A)}{l_B}_{\mathbf{m},N-N_A}|^2, 
\end{align}
with
\begin{align}
\bra{l_B}_{\mathbf{m},N-N_A}\ket{\Psi(N_A,k_A)}=\frac{1}{\sqrt{2^{N-N_A}}}\sum_{k_B=0}^{N-N_A}\sqrt{\binom{N-N_A}{k_B}}e^{- i \frac{\mu}{2} (N/2 - k_A - k_B)^2}\bra{l_B}_{\vec{m},N-N_A}\ket{k_B}_{N-N_A}.
\end{align}

\section{V.\quad Other results}

As depicted schematically in Figs.~\ref{fig:SIcriteria} and~\ref{fig:SIfirstterm} below, we also investigate the optimized criteria and their respective first terms (expressing the assisted phase sensitivity) for split spin states with different total atom numbers $N$, where $N$ is assumed to be even.

\begin{figure*}[h]
	\begin{center}
		\includegraphics[width=\textwidth]{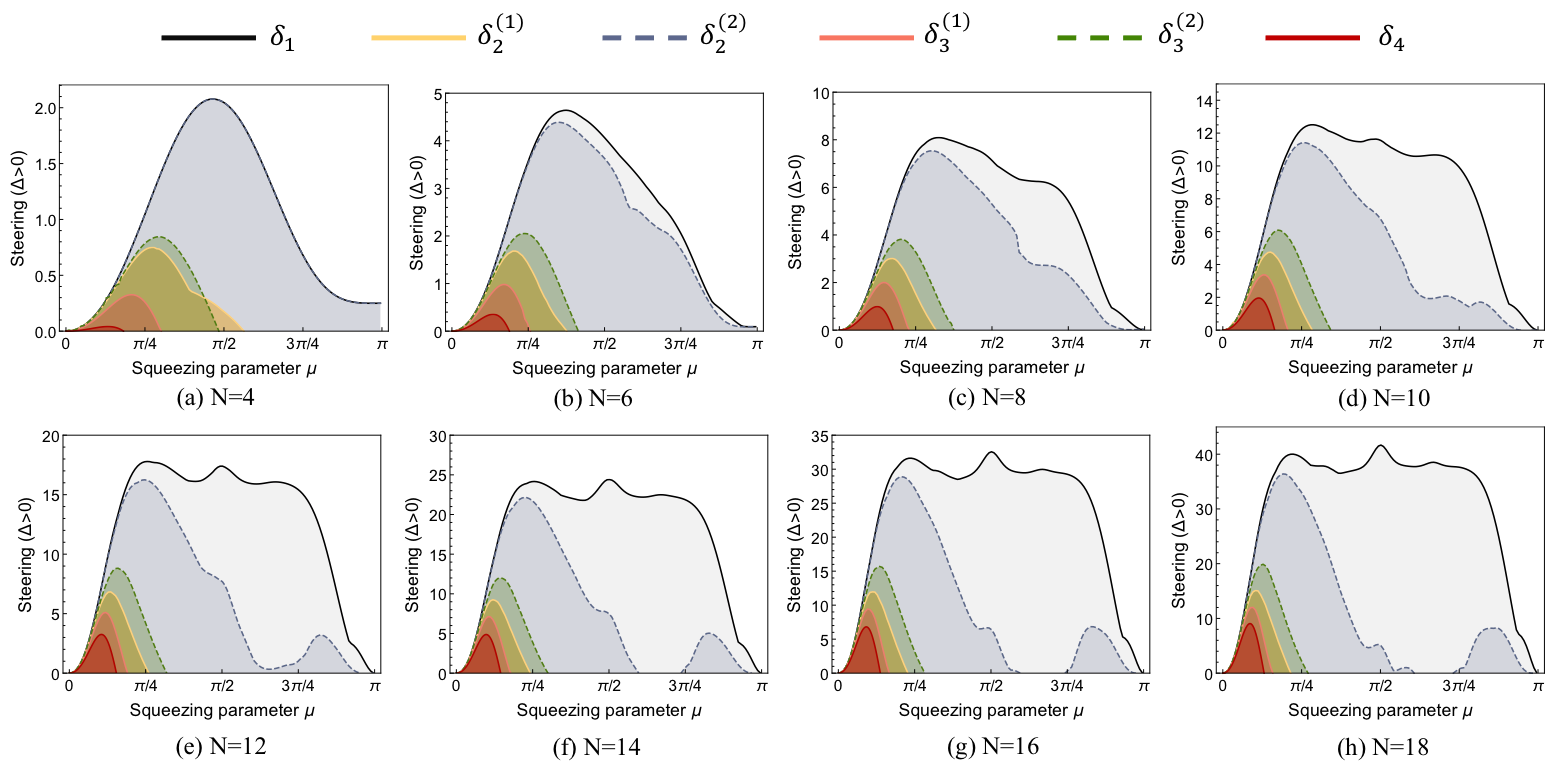}
	\end{center}
	\caption{Optimized criterion for split spin states $\ket{\Phi(\mu)}$ with a total number of atoms $N=4,6,8,...,18$. The hierarchy  (\ref{eq:hierarchy1}) and (\ref{eq:hierarchy2}) are clearly illustrated here. In addition, when the atom number is small ($N \leq 8$), our second-order conditional spin-squeezing paramter criterion $\delta_2^{(2)}$ is able to reveal steering in the same range as the Fisher criterion $\delta_1$ does, and even coincides with $\delta_1$ when $N=4$.}
	\label{fig:SIcriteria}
\end{figure*}
\begin{figure*}[t]
	\begin{center}
		\includegraphics[width=\textwidth]{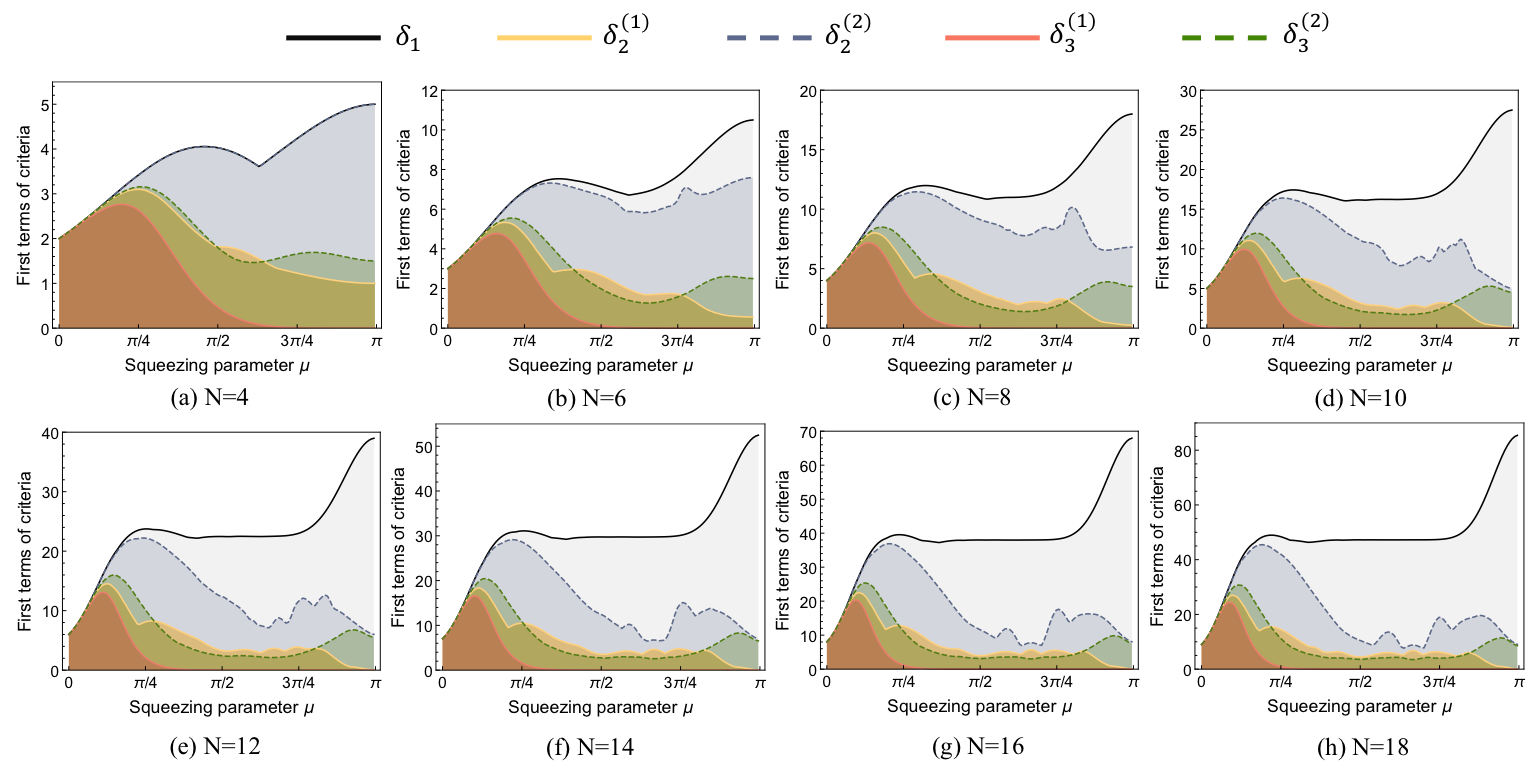}
	\end{center}
	\caption{First terms of optimized criteria $\delta_{1,2,3}$, i.e., the chain of inequalities~(\ref{eq:lwrbndmethods}), for split spin states $\ket{\Phi(\mu)}$ with a total number of atoms $N=4,6,8,...18$. It can been seen that when $N=4$, the first term of $\delta_2^{(2)}$ is equal to $\delta_1$, leading to the agreement of the two criteria in this case.}
	\label{fig:SIfirstterm}
\end{figure*}

\end{widetext}


\begin{thebibliography}{}


\bibitem{Schrodinger1935} E. Schr\"{o}dinger, Discussion of probability relations between separated systems \href{https://doi.org/10.1017/S0305004100013554}{Proc. Cambridge Philos. Soc. \textbf{31}, 555 (1935)}.

\bibitem{EPR1935} A. Einstein, B. Podolsky, and N. Rosen, Can Quantum-Mechanical Description of Physical Reality Be Considered Complete? \href{https://doi.org/10.1103/PhysRev.47.777}{Phys. Rev. \textbf{47}, 777 (1935)}.

\bibitem{WisemanPRL2007} H. M. Wiseman, S. J. Jones, and A. C. Doherty, Steering, Entanglement, Nonlocality, and the Einstein-Podolsky-Rosen Paradox, \href{https://doi.org/10.1103/PhysRevLett.98.140402}{Phys. Rev. Lett. \textbf{98}, 140402 (2007)}.

\bibitem{ReidRMP2009} M. D. Reid, P. D. Drummond, W. P. Bowen, E. G. Cavalcanti, P. K. Lam, H. A. Bachor, U. L. Andersen, and G. Leuchs, Colloquium: The Einstein-Podolsky-Rosen paradox: From concepts to applications, \href{https://doi.org/10.1103/RevModPhys.81.1727}{Rev. Mod. Phys. \textbf{81}, 1727 (2009)}.

\bibitem{CavalcantiRPP2017} D. Cavalcanti and P. Skrzypczyk, Quantum steering: A review with focus on semidefinite programming, \href{https://iopscience.iop.org/article/10.1088/1361-6633/80/2/024001}{Rep. Prog. Phys. \textbf{80}, 024001 (2017)}.

\bibitem{GuehneRMP2020} R. Uola, A. C. S. Costa, H. C. Nguyen, and O. G\"{u}hne, Quantum steering, \href{https://doi.org/10.1103/RevModPhys.92.015001}{Rev. Mod. Phys. \textbf{92}, 015001 (2020)}.

\bibitem{QiongyiPRL2015} Q. He, L. Rosales-Z\'{a}rate, G. Adesso, and M. D. Reid, Secure Continuous Variable Teleportation and Einstein-Podolsky- Rosen Steering, \href{https://doi.org/10.1103/PhysRevLett.115.180502}{Phys. Rev. Lett. \textbf{115}, 180502 (2015)}.

\bibitem{ChiuNPJQI2016} C.-Y. Chiu, N. Lambert, T.-L. Liao, F. Nori, and C.-M. Li, No-cloning of quantum steering,  \href{https://www.nature.com/articles/npjqi201620}{npj Quantum Inf. \textbf{2}, 16020 (2016)}.

\bibitem{BranciardPRA2012} C. Branciard, E. G. Cavalcanti, S. P. Walborn, V. Scarani, and
H. M. Wiseman, One-sided device-independent quantum key distribution: Security, feasibility, and the connection with steering,  \href{https://doi.org/10.1103/PhysRevA.85.010301}{Phys. Rev. A \textbf{85}, 010301(R) (2012)}.

\bibitem{GehringNC2015} T. Gehring, V. H\"{a}ndchen, J. Duhme, F. Furrer, T. Franz, C. Pacher, R. F. Werner, and R. Schnabel, Implementation of continuous-variable quantum key distribution with compos- able and one-sided-device-independent security against coherent attacks,  \href{https://www.nature.com/articles/ncomms9795}{Nat. Commun. \textbf{6}, 8795 (2015)}.

\bibitem{WalkOpt2016} N. Walk, S. Hosseini, J. Geng, O. Thearle, J. Y. Haw, S. Armstrong, S. M. Assad, J. Janousek, T. C. Ralph, T. Symul, H. M. Wiseman, and P. K. Lam, Experimental demonstration of Gaussian protocols for one-sided device-independent quantum key distribution,  \href{https://doi.org/10.1364/OPTICA.3.000634}{Optica \textbf{3}, 634 (2016)}.

\bibitem{GallegoPRX2015} R. Gallego and L. Aolita, Resource Theory of Steering, 
 \href{https://doi.org/10.1103/PhysRevX.5.041008}{Phys. Rev. X \textbf{5}, 041008 (2015)}.

\bibitem{ArmstrongNP2015} S. Armstrong, M. Wang, R. Y. Teh, Q. Gong, Q. He, J. Janousek, H. A. Bachor, M. D. Reid, and P. K. Lam, Multipartite Einstein-Podolsky-Rosen steering and genuine tripartite entanglement with optical networks, \href{https://www.nature.com/articles/nphys3202}{Nat. Phys. \textbf{11}, 167 (2015)}.

\bibitem{XiangPRA2017} Y. Xiang, I. Kogias, G. Adesso, and Q. He, Multipartite Gaussian steering: Monogamy constraints and quantum cryptography applications, \href{https://doi.org/10.1103/PhysRevA.95.010101}{Phys. Rev. A \textbf{95}, 010101(R) (2017)}.

\bibitem{KogiasPRA2017} I. Kogias, Y. Xiang, Q. He, and G. Adesso, Unconditional security of entanglement-based continuous-variable quantum secret sharing, \href{https://doi.org/10.1103/PhysRevA.95.012315}{Phys. Rev. A \textbf{95}, 012315 (2017)}.

\bibitem{YadinNC2021} B. Yadin, M. Fadel, and M. Gessner, Metrological complementarity reveals the Einstein-Podolsky-Rosen paradox, \href{https://doi.org/10.1038/s41467-021-22353-3 }{Nat. Commun. \textbf{12}, 2401 (2021)}.

\bibitem{ReidPRA1989} M. D. Reid, Demonstration of the Einstein-Podolsky-Rosen paradox using nondegenerate parametric amplification, \href{https://doi.org/10.1103/PhysRevA.40.913}{Phys. Rev. A \textbf{40}, 913 (1989)}.

\bibitem{GuoPRA2019} Y. Guo, W. Ye, H. Zhong, and Q. Liao, Continuous-variable quantum key distribution with non-Gaussian quantum catalysis, \href{ https://doi.org/10.1103/PhysRevA.99.032327}{Phys. Rev. A \textbf{99}, 032327 (2019)}.

\bibitem{TakahashiNP2010} H. Takahashi, J. S. Neergaard-Nielsen, M. Takeuchi, M. 
Takeoka, K. Hayasaka, A. Furusawa, and M. Sasaki, Entanglement distillation from Gaussian input states, \href{https://www.nature.com/articles/nphoton.2010.1}{Nat. photonics \textbf{4}, 178-181 (2010)}.

\bibitem{LeeNPJQI2019} J. Lee, J. Park, and H. Nha, Quantum non-Gaussianity and secure quantum communication, \href{https://doi.org/10.1038/s41534-019-0164-9}{npj Quantum Information \textbf{5}, 49 (2019)}.

\bibitem{MariPRL2012} A. Mari and J. Eisert, Positive Wigner functions render classical simulation of quantum computation efficient, \href{https://doi.org/10.1103/PhysRevLett.109.230503}{Phys. Rev. Lett. \textbf{109}, 230503 (2012)}.

\bibitem{WalbornPRL11}
S. P. Walborn, A. Salles, R. M. Gomes, F. Toscano, and P. H. Souto Ribeiro, Revealing Hidden Einstein-Podolsky-Rosen Nonlocality, \href{https://doi.org/10.1103/PhysRevLett.106.130402}{Phys. Rev. Lett. \textbf{106}, 130402 (2013)}.

\bibitem{WalbornPRA13} J. Schneeloch, C. J. Broadbent, S. P. Walborn, E. G. Cavalcanti and J. C. Howell, Einstein-Podolsky-Rosen steering inequalities from entropic uncertainty relations, \href{https://doi.org/10.1103/PhysRevA.87.062103}{Phys. Rev. A \textbf{87}, 062103 (2013)}.




\bibitem{ShenPRL2015} Y. Shen, S. M. Assad, N. B. Grosse, X. Y. Li, M. D. Reid, and P. K. Lam, Nonlinear Entanglement and its Application to Generating Cat States, \href{https://doi.org/10.1103/PhysRevLett.114.100403}{Phys. Rev. Lett. \textbf{114}, 100403 (2015)}.

\bibitem{HilleryPRL2006} M. Hillery and M. S. Zubairy, Entanglement conditions for two-mode states, \href{https://doi.org/10.1103/PhysRevLett.96.050503}{Phys. Rev. Lett. \textbf{96}, 050503 (2006)}.

\bibitem{CavalcantiPRA2011} E. G. Cavalcanti, Q. Y. He, M. D. Reid, and H. M. Wiseman, Unified criteria for multipartite quantum nonlocality, \href{https://doi.org/10.1103/PhysRevA.84.032115}{Phys. Rev. A \textbf{84}, 032115 (2011)}.

\bibitem{QiongyiPRA2012} Q. Y. He, P. D. Drummond, M. K. Olsen, and M. D. Reid, Einstein-Podolsky-Rosen entanglement and steering in two-well Bose-Einstein-condensate ground states, \href{https://doi.org/10.1103/PhysRevA.86.023626}{Phys. Rev. A \textbf{86}, 023626 (2012)}.

\bibitem{PezzeRMP2018} L. Pezz$\grave{\text{e}}$, A. Smerzi, M. K. Oberthaler, R. Schmied, and P. Treutlein, Quantum metrology with nonclassical states of atomic ensembles, \href{ https://doi.org/10.1103/RevModPhys.90.035005}{Rev. Mod. Phys. \textbf{90}, 035005 (2018)}.

\bibitem{ManuelPRL2019} M. Gessner, A. Smerzi, and Luca Pezze, Metrological Nonlinear Squeezing Parameter, \href{https://doi.org/10.1103/PhysRevLett.122.090503}{Phys. Rev. Lett. \textbf{122}, 090503 (2019)}.

\bibitem{BaamaraArxiv} Y. Baamara, A. Sinatra, M. Gessner, Scaling laws for the sensitivity enhancement of non-Gaussian spin states, \href{https://arxiv.org/abs/2105.11421v1}{arXiv:2105.11421}.

\bibitem{StrobelScience2014} H. Strobel, W. Muessel, D. Linnemann, T. Zibold, D. B. Hume, L. Pezze`, A. Smerzi, and M. K. Oberthaler, Fisher information and entanglement of non-Gaussian spin states, \href{https://science.sciencemag.org/content/345/6195/424}{Science \textbf{345}, 424 (2014)}.

\bibitem{BohnetSCIENCE2016} J. G. Bohnet, B. C. Sawyer, J. W. Britton, M. L. Wall, A. M. Rey, M. Foss-Feig, J. J. Bollinger, Quantum spin dynamics and entanglement generation with hundreds of trapped ions, \href{https://doi.org/10.1126/science.aad9958 }{Science \textbf{352}, 1297 (2016)}.

\bibitem{EvrardPRL2019} A. Evrard, V. Makhalov, T. Chalopin, L. A. Sidorenkov, J. Dalibard, R. Lopes, and S. Nascimbene, Enhanced Magnetic Sensitivity with Non-Gaussian Quantum Fluctuations, \href{https://doi.org/10.1103/PhysRevLett.122.173601}{Phys. Rev. Lett. \textbf{122}, 173601 (2019)}.

\bibitem{XuArXiv2021} K. Xu, Y.-R. Zhang, Z.-H. Sun, H. Li, P. Song, Z. Xiang, K. Huang, H. Li, Y.-H. Shi, C.-T. Chen, X. Song, D. Zheng, F. Nori, H. Wang, H. Fan, Metrological characterisation of non-Gaussian entangled states of superconducting qubits, \href{https://arxiv.org/abs/2103.11434}{arXiv:2103.11434}.

\bibitem{ColomboArXiv2021} S. Colombo, E. Pedrozo-Pe\~{n}afiel, A. F. Adiyatullin, Z. Li, E. Mendez, C. Shu, V. Vuletic, Time-Reversal-Based Quantum Metrology with Many-Body Entangled States, \href{https://arxiv.org/abs/2106.03754}{arXiv:2106.03754}.

\bibitem{WinelandPRA1992} D. J. Wineland, J. J. Bollinger, W. M. Itano, F. L. Moore, and D. J. Heinzen, Spin squeezing and reduced quantum noise in spectroscopy, \href{ https://doi.org/10.1103/PhysRevA.46.R6797}{Phys. Rev. A \textbf{46}, R6797 (1992)}.

\bibitem{SorensenNL2001} A. S\o rensen, L. M. Duan, J. I. Cirac, and P. Zoller, Many-particle entanglement with Bose-Einstein condensates, \href{https://www.nature.com/articles/35051038}{Nature(London) \textbf{409}, 63 (2001)}.

\bibitem{SorensenPRL2001} A. S. S\o rensen and K. M\o lmer, Entanglement and Extreme Spin Squeezing, \href{https://doi.org/10.1103/PhysRevLett.86.4431}{Phys. Rev. Lett. \textbf{86}, 4431 (2001)}.

\bibitem{PezzePRL2009} L. Pezz\'{e} and A. Smerzi, Entanglement, Nonlinear Dynamics, and the Heisenberg Limit, \href{https://doi.org/10.1103/PhysRevLett.102.100401}{Phys. Rev. Lett. \textbf{102}, 100401 (2009)}.

\bibitem{RenPRL2021} Z. Ren, W. Li, A. Smerzi, and M. Gessner, Metrological Detection of Multipartite Entanglement from Young Diagrams, \href{https://doi.org/10.1103/PhysRevLett.126.080502}{Phys. Rev. Lett. \textbf{126}, 080502 (2021)}.



\bibitem{Holevo1982} A. S. Holevo, Probabilistic and Statistical Aspects of Quantum Theory (North-Holland, Amsterdam, 1982). 

\bibitem{Helstrom1976} C. W. Helstrom, Quantum Detection and Estimation Theory (Academic Press, New York, 1976). 

\bibitem{ParisIJQI2009} M. G. A. Paris, Quantum Estimation for Quantum Technology, 
\href{https://doi.org/10.1142/S0219749909004839}{Int. J. Quant. Inf. \textbf{07}, 125-137 (2009)}.


\bibitem{BraunsteinPRL1994} S. L. Braunstein and C. M. Caves, Statistical Distance and the Geometry of Quantum States, \href{https://doi.org/10.1103/PhysRevLett.72.3439}{Phys. Rev. Lett. \textbf{72}, 3439 (1994)}.



\bibitem{footnote} Note that in general the optimal $M$ depends on the measurement setting and result.

\bibitem{BraunsteinAnnPhys1996} S. L. Braunstein, C. M. Caves, G. J. Milburn, Generalized Uncertainty Relations: Theory, Examples, and Lorentz Invariance, \href{https://doi.org/10.1006/aphy.1996.0040}{Ann. Phys.
\textbf{247}, 135 (1996)}.

\bibitem{CavesPRD1981} C. M. Caves, Quantum-mechanical noise in an interferometer, \href{ https://doi.org/10.1103/PhysRevD.23.1693}{Phys. Rev. D \textbf{23}, 1693 (1981)}.

\bibitem{TothJPA2014} G. T\'{o}th, and I. Apellaniz, Quantum metrology from a quantum information science perspective, \href{https://iopscience.iop.org/article/10.1088/1751-8113/47/42/424006}{J. Phys. A \textbf{47}, 424006 (2014)}.

\bibitem{MaPR2011} J. Ma, X. Wang, C. Sun, and F. Nori, Quantum spin squeezing, \href{https://linkinghub.elsevier.com/retrieve/pii/S0370157311002201}{Phys. Rep. \textbf{509}, 89 (2011)}.

\bibitem{LudlowRMP2015} A. D. Ludlow, M. M. Boyd, J. Ye, E. Peik, and P. O. Schmidt, Optical atomic clocks, \href{ https://doi.org/10.1103/RevModPhys.87.637}{Rev. Mod. Phys. \textbf{87}, 637 (2015)}.

\bibitem{GodaNP2008} K. Goda, O. Miyakawa, E. E. Mikhailov, S. Saraf, R. Adhikari, K. McKenzie, R. Ward, S. Vass, A. J. Weinstein, and N. Mavalvala, A quantum-enhanced prototype gravitational-wave detector, \href{https://www.nature.com/articles/nphys920}{Nature Phys. \textbf{4}, 472-476 (2008)}.

\bibitem{RiedelNL2010} M. F. Riedel, P. B\"{o}hi, Y. Li, T. W. H\"{a}nsch, A. Sinatra, and P. Treutlein, Atom-chip-based generation of entanglement for quantum metrology, \href{https://www.nature.com/articles/nature08988}{Nature (London) \textbf{464}, 1170 (2010)}.


\bibitem{LerouxPRL2010} I. D. Leroux, M. H. Schleier-Smith, and V. Vuleti\'{c}, Implementation of Cavity Squeezing of a Collective Atomic Spin, \href{https://doi.org/10.1103/PhysRevLett.104.073602}{Phys. Rev. Lett. \textbf{104}, 073602 (2010)}.

\bibitem{MonzPRL2011} T. Monz, P. Schindler, J. T. Barreiro, M. Chwalla, D. Nigg, W. A. Coish, M. Harlander, W. H\"{a}nsel, M. Hennrich, and R. Blatt, 14-Qubit Entanglement: Creation and Coherence,  \href{ https://doi.org/10.1103/PhysRevLett.106.130506}{Phys. Rev. Lett. \textbf{106}, 130506 (2011)}.

\bibitem{ChalopinNC2018} T. Chalopin, C. Bouazza, A. Evrard, V. Makhalov, D. Dreon, J. Dalibard, L. A. Sidorenkov, and S. Nascimbene, Quantum-enhanced sensing using non-classical spin states of a highly magnetic atom, \href{https://doi.org/10.1038/s41467- 018-07433-1}{Nat. Commun. \textbf{9}, 4955 (2018)}.

\bibitem{KitagawaPRA1993} M. Kitagawa and M. Ueda, Squeezed spin states, \href{ https://doi.org/10.1103/PhysRevA.47.5138}{Phys. Rev. A \textbf{47}, 5138 (1993)}.

\bibitem{YumangNJP2019} Y. Jing, M. Fadel, V. Ivannikov, and T. Byrnes, Split spin-squeezed Bose-Einstein condensates, \href{https://doi.org/10.1088/1367-2630/ab3fcf }{New J. Phys. \textbf{21}, 093038 (2019)}.

\bibitem{FadelScience2018} M. Fadel, T. Zibold, B. D\'{e}camps, and P. Treutlein, Spatial entanglement patterns and Einstein-Podolsky-Rosen steering in Bose-Einstein condensates, \href{https://science.sciencemag.org/content/360/6387/409}{Science \textbf{360}, 409-413 (2018)}.














\end{thebibliography}
\end{document}